\documentclass[conference,screen]{IEEEtran}
\usepackage[noadjust]{cite}

\usepackage{framed}
\usepackage{multirow}
\usepackage{booktabs}
\usepackage{ifthen}
\usepackage{color}
\usepackage{url}

\usepackage[T1]{fontenc}
\usepackage{xspace}
\usepackage{graphicx}
\usepackage{subfigure}
\usepackage{amsfonts}
\usepackage{amsmath}
\usepackage{hyperref,endnotes}
\usepackage{breakurl}
\usepackage[misc]{ifsym}
\usepackage{listings}

\definecolor{pblue}{rgb}{0.13,0.13,1}
\definecolor{pgreen}{rgb}{0,0.5,0}
\definecolor{pred}{rgb}{0.9,0,0}
\definecolor{pgrey}{rgb}{0.46,0.45,0.48}

\begin{document}

\title{A Deep Dive into NFT Rug Pulls}

\author{\IEEEauthorblockN{Jintao Huang\IEEEauthorrefmark{1}, Ningyu He\IEEEauthorrefmark{2}, Kai Ma\IEEEauthorrefmark{1}, Jiang Xiao\IEEEauthorrefmark{1}, and Haoyu Wang\IEEEauthorrefmark{1}}
\IEEEauthorblockA{\IEEEauthorrefmark{1}\textit{Huazhong University of Science and Technology}\\\IEEEauthorrefmark{2}\textit{Peking University}}
}

\maketitle

\begin{abstract}

NFT rug pull is one of the most prominent type of scam that the developers of a project abandon it and then run away with investors' funds. Although they have drawn attention from our community, to the best of our knowledge, the NFT rug pulls have not been systematically explored. To fill the void, this paper presents the first in-depth study of NFT rug pulls. Specifically, we first compile a list of 253 known NFT rug pulls as our initial ground truth, based on which we perform a pilot study, highlighting the key symptoms of NFT rug pulls. Then, we enforce a strict rule-based method to flag more rug pulled NFT projects in the wild, and have labelled 7,487 NFT rug pulls as our extended ground truth. Atop it, we have investigated the art of NFT rug pulls, with kinds of tricks including explicit ones that are embedded with backdoors, and implicit ones that manipulate the market. To release the expansion of the scam, we further design a prediction model to proactively identify the potential rug pull projects in an
early stage ahead of the scam happens. We have implemented a prototype system deployed in the real-world setting for over 5 months. Our system has raised alarms for 7,821 NFT projects, by the time of this writing, which can work as a whistle blower
that pinpoints rug pull scams timely, thus mitigating the impacts.

\end{abstract}
\section{Introduction}
\label{sec:intro}

NFTs, or non-fungible tokens, were introduced as a new type of cryptocurrency in October 2015 with the launch of the first NFT project \texttt{Etheria}. Since then, NFTs have attracted significant attention from the public and opened up new possibilities. 
From 2020 to 2021, the NFT market has experienced explosive growth, totally an increase of \$17 billion, which is a staggering 210X compared to 2020's total of \$82 million~\cite{nftwikipedia}. 
Several NFT collections, e.g., \texttt{Mutant Ape Yacht Club}, \texttt{Azuki}, and \texttt{Bored Aple Yacht Club}, have generated significant sales volumes in NFT markets, with values reaching up to millions of USD~\cite{OpenSeatop}. 
One of the key features of NFTs is their \textit{non-fungible} nature, which enables them to be linked to a specific digital asset, such as images, art, music, and sports highlights. 
This association may confer licensing rights to use the asset for a specified purpose. 
To enable the implementation of NFTs, Ethereum proposes two standards: \texttt{ERC-721} and \texttt{ERC-1155}.
The former generates tokens that are \textit{one-of-a-kind} and linked to a unique token ID, while the latter generates a set of tokens that may share the same ID.

Despite their popularity, NFTs, like other types of cryptocurrencies, are prone to security threats. Attackers and scammers are staring at the huge market, which are frequently reported in the media outlets~\cite{hackexample1, hackexample2}. 
Among them, \textit{rug pull}~\cite{rugpull} is one of the most prominent type of scam, i.e., developers of a project abandon it and run away with investors’ funds. According the research of \texttt{Chainanalysis}, cryptocurrency investors in 2021 lost over \$2.8 billion to rug pulls, and NFT rug pull is on the rise~\cite{chainanalysis}.
For example, the \texttt{Frosties} is a famous NFT rug pull scam, which led to the theft of over \$1.3 million, after which the two founders were charged for the maximum sentence of 20 years in prison~\cite{FrostiesNFT}.

NFT rug pulls have already drawn attention from our community. For example, there are several crowd-sourcing based channels maintaining a list NFT rug pull scams based on manual effort~\cite{rugpullfinder,chainabuse}. Despite this, to the best of our knowledge, the NFT rug pull scams have not been systematically investigated or measured. Thus, there is a general \textit{lack of an understanding} of NFT rug pulls, including 1) the concrete patterns of NFT rug pulls, 2) to what extent NFT rug pulls exist in the ecosystem, 3) the sophisticated tricks exploited by them, and more importantly, 4) no existing approaches can be used to detect, mitigate or prevent this kind of scam. 

\textbf{This work.} 
In this paper, we present the first systematic study of NFT rug pulls. To understand the concrete symptoms of NFT rug pulls, we first harvest a list of 253 known NFT rug pulls (i.e., \textit{initial ground truth}) revealed by our community, and based on which to perform a pilot study, highlighting the key features related to NFT rug pulls (\S\ref{sec:symptom}). Then, we design an effective detector to measure the prevalence of NFT rug pulls in the ecosystem, by analyzing all the on-chain and off-chain data related to over 173K Ethereum NFT projects (\S\ref{sec:prevalence}). Our detector has labelled 7,487 rug pulled NFT projects (i.e., \textit{extended ground truth}) that have taken place in the ecosystem, $30\times$ greater than existing crowd-sourcing based collections. 
Then, we investigate the most popular tricks used in these scams (\S\ref{sec:tricks}), e.g., gaining a profit by taking advantage of backdoors in smart contracts or manipulating markets. 
We have observed that, beyond the pump and dump nature of NFT rug pull scams, 84\% of the scam projects have exploited at least one kind of tricks to facilitate the delivery of the scam. 
At last, to explore whether we can raise alarms of NFT rug pull projects before the scam happens, we have created a prediction model taking advantage of 73 kinds of features extracted from both on-chain and off-chain data (\S\ref{sec:predicting}). Our model can raise alarms for 90\% of NFT scam events within 96 hours ahead of rug pull happens. We have implemented a prototype system to monitor NFT rug pulls in the wild since November 2022. By the time of this writing, we have successfully raise warnings for 7,821 NFT projects (beyond the extended ground truth of 7,487 cases), and most of them have been confirmed to be rug pulls in later times with additional evidences.

We make the following main research contributions:
\begin{itemize}
    \item We have uncovered the key symptoms of NFT rug pulls and devised an effective approach to pinpoint NFT rug pulls that have already taken place in the wild. 
    We have flagged 7,487 rug pulled NFT projects in total using a strict rule-based approach, \textit{by far the largest NFT rug pull dataset}, 30 times greater than existing efforts. This can be served as the reliable benchmark for future research in our community.

    \item Except for the traditional pump and dump scam, we have revealed eight kinds of widely used tricks in NFT rug pulls, which are categorized into two categories, i.e., explicit rug pulls that taking advantage of the backdoors hidden in the smart contract, and implicit rug pulls that manipulating markets. We further design analyzers to automatically label these tricks, and observe 84\% of existing rug pulls show such behaviors.

    \item We have proposed a real-time solution to mitigate the impact of NFT rug pulls. Specifically, we design a prediction model to raise alarms ahead of the scam happens, based on the signals (e.g., token transfers and trade events) released in their early stages. It can work as a \textit{whistle blower} that pinpoints rug pull scams timely, thus mitigating the impacts.

    \item We have implemented a prototype NFT rug pull warning system, which has been deployed on Ethereum to monitor NFT transactions since November 2022. By the time of this writing, we have successfully raise alarms for 7,821 new NFT rug pull projects (beyond the 7,487 ground truth). Although it is hard to verify all of them, we make great efforts to show that most of them are indeed rug pulls with additional evidences, while we can raise warnings several days earlier.
    
\end{itemize}

All our dataset will be released to the research community.
\section{Background}
\label{sec:background}
In this section, we will introduce the necessary background of NFTs and the rug pull scams. Further, we give a concrete example to depict the general process of NFT rug pull scam.

\subsection{Ethereum \& Non-Fungible Token (NFT)}
\label{sec:background:eth}
Along with the prosperity of Bitcoin~\cite{Bitcoin}, Ethereum~\cite{Ethereum} occupies the second leading position among blockchain platforms, on which developers can deploy \textit{smart contracts} that can be automatically executed if the condition is met. Smart contracts can interact with each other, while the interactions are recorded on-chain and is accessible for anyone, which is called \textit{transactions}.
Furthermore, transactions can not only send data across smart contracts, but also transfer tokens. For example, the official token in Ethereum is \textit{Ether} (i.e., ETH), which is a type of cryptocurrency that can be circulated in exchanges.
Within certain transactions, \textit{events} will be emitted and recorded on-chain to provide additional details about the corresponding transaction, e.g., initiator and carrying data.

Except for official tokens, accounts (i.e., users) in Ethereum can issue tokens following the required standards. An ``Ethereum Request for Comments'' (ERC)~\cite{erc} is a standard protocol that is used in issuing tokens under the Ethereum blockchain. Specifically, ERC-20~\cite{erc20} is the most widely adopted standard. It has six interfaces that should be implemented by developers, e.g., \texttt{transfer}, which allows the initiator to transfer a certain amount of current ERC-20 token to someone else. Note that, Ethereum only examines whether interfaces are implemented instead of the correctness of implementations.
\texttt{ERC-721}~\cite{erc721} and \texttt{ERC-1155}~\cite{erc1155} are two emerging standards for NFTs. Tokens under these two standards have a specific characteristic, named \textit{non-fungibility}.
Specifically, tokens under \texttt{ERC-721} are one-of-a-kind, i.e., each token corresponds to a unique ID. Tokens under \texttt{ERC-1155} also possess this characteristic, while an ID can be linked to several indistinguishable tokens.
To this end, the distinguishability among \textit{non-fungible tokens} (NFTs) can be used to bind real-world unreplicable items, such as an artwork or a piece of music~\cite{NFTsForArtists}, in one-to-one relationships. 
NFTs are becoming popular especially in the art world as a means for artists to monetize and sell their digital creations~\cite{beeplehomepage}. 
There are many tools that can assist in binding artworks with newly minted NFTs, e.g., OpenSea~\cite{opensea_2022_opensea} or Rarible~\cite{rarible}.
Note that, NFTs can also be traded in secondary markets~\cite{opensea_2022_opensea} (i.e., where investors buy and sell securities from other investors), thus the value of an NFT is highly determined by its supply and demand relationship, or its popularity in the secondary market. 
After a successful trading, the ownership of the NFT will be transferred from the seller to the buyer, through the smart contract of the market. 
The lowest price of trade within recent period is defined as the \textit{floor price}, which is a vital indicator to measure the value of an NFT project.

\subsection{Rug Pull}
\label{sec:background:rugpull}
In the blockchain domain, \textit{rug pull} is a type of scams that developers of a project abandon it and run away with investors’ funds.
The rug pull scams have been widely found in
\textit{decentralized exchanges} (\textit{DEXs}) where the ultimate purpose is to fool the victims to invest in the scam tokens they created and then drain the money of the pools~\cite{xia2021trade}.
As a general type of scams, similar tricks are emerging in the NFT ecosystem. Specially, NFT rug pulls are divided into three stages.

\begin{enumerate}
	\item \textbf{Luring victims.} At first, scammers will lure victims by promoting their projects on social media platforms, e.g., \texttt{Twitter} and \texttt{Discord}. Meanwhile, they use kinds of tricks (e.g., spread fake advertisements or post fake comments using bots) to make an illusion that the NFTs they promoted will be in great demand, and investing in them would reap high rewards.
	\item \textbf{Pumping up the price and making a profit.} To pursue maximum profits, these rug pullers will try their best to make their projects \textit{look} valuable and attractive, e.g., by fabricating trading volumes. After the price has been driven up, they will sell their owned NFTs, and pull as much value out of them as possible. Even some NFT contracts have backdoors that the scammers can over-mint NFTs to earn extra profit (\S\ref{sec:tricks}).
	\item \textbf{Running away.} Finally, the rug pullers will run away with investors' fund, abandon the project, and usually deactivate their social media. Such a project with no endorsements will not attract any investors or collectors, leading to an extreme low price on secondary markets. All the NFT holders under this project are victims that suffer financial losses.
 
\end{enumerate}

\subsection{Example: AniMoon Rug Pull}
\label{sec:background:example}
\begin{figure}[t]
    \centering
    \includegraphics[width=\columnwidth]{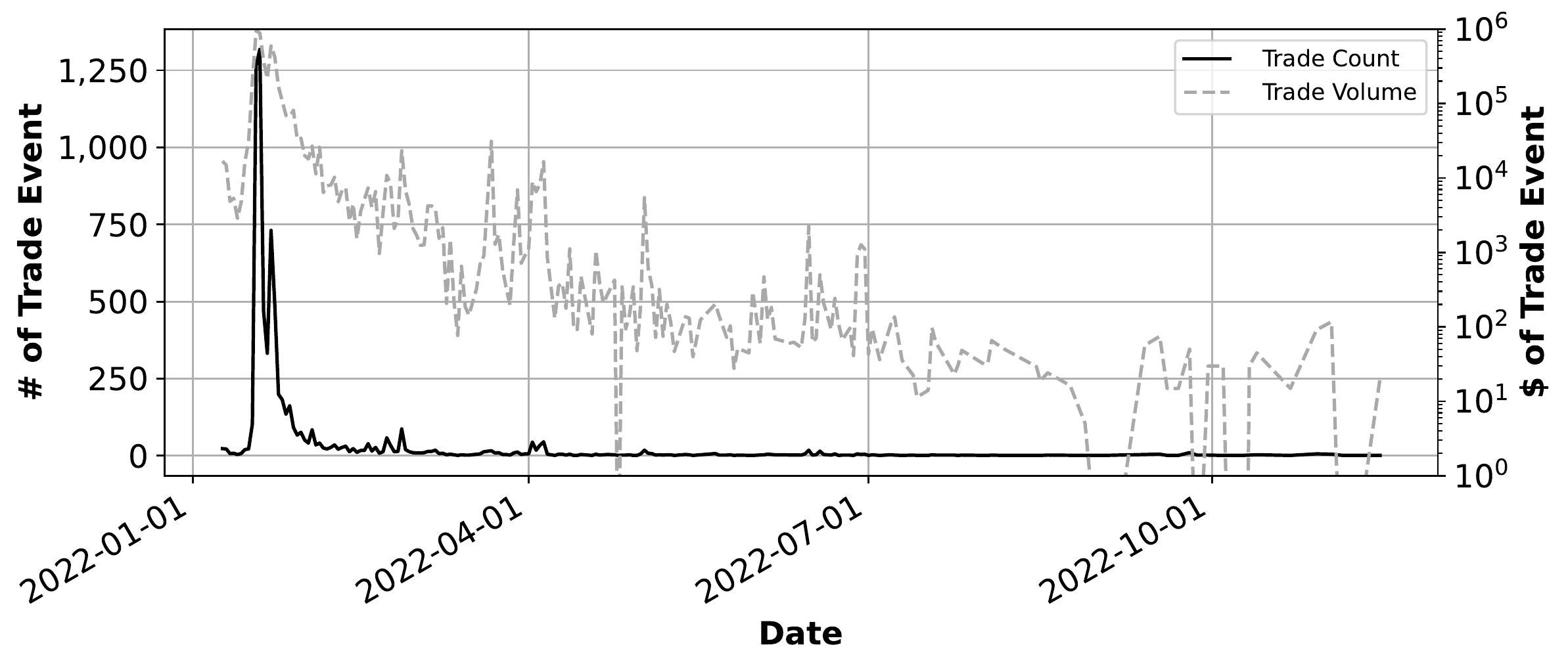}
    \caption{The activity of AniMoon on OpenSea. }
    % \hny{This figure should be revised. I cannot distinguish the two lines.}}\jt{fix}
    \label{fig:animoon}
\end{figure}

\begin{figure}[t]
    \centering
    \includegraphics[width=0.9\columnwidth]{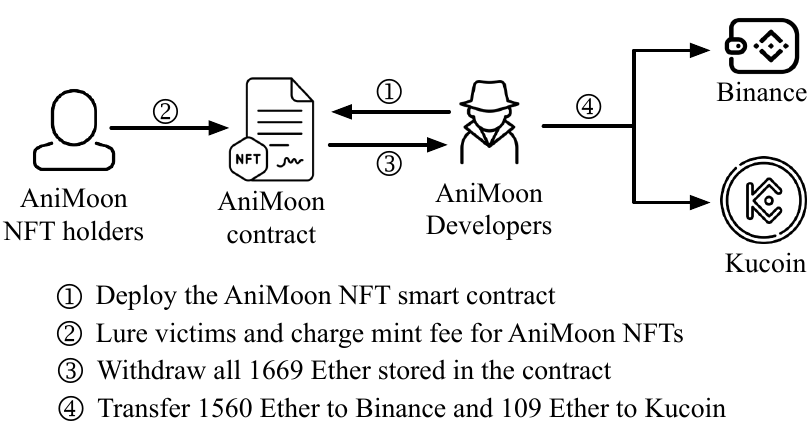}
    \caption{The whole process of AniMoon rug pull.}
    \label{fig:animoonprocess}
\end{figure}

\texttt{AniMoon} is a play-to-earn (P2E) game based on the famous \texttt{Nintendo Pokémon} cartoon series. Fig.~\ref{fig:animoon} shows its transaction numbers and volumes in the largest NFT secondary market, \texttt{OpenSea}, and Fig.~\ref{fig:animoonprocess} illustrates the whole process of how the rug pull is carried.
We will briefly introduce its three stages as we introduced in \S\ref{sec:background:rugpull}.

\subsubsection{Luring Victims}
The team attracted unsuspecting players by advertising on various social media platforms. For example, they have launched an official website~\cite{AnimoonOfficial} and created a \texttt{Discord} server~\cite{AnimoonDiscord}. 
Two founders also advertised the project on their own \texttt{Twitter} accounts~\cite{AnimoonTwitteAcount1, AnimoonTwitteAcount2}, with 94.4K and 1.4K followers, respectively.
The massive user base has successfully attracted more than 9K users (i.e., victims) to invest in the project. According to our statistics, more than 90K users are invited into their \texttt{Discord} server.
As incentives, the owner promised physical rewards and cash dividends.
Such a seemingly low-risk but high-returns project finally lures a large number of victims.

\subsubsection{Pumping up the price and making a profit}
As illustrated in Fig.~\ref{fig:animoon}, the invested money takes a rocket. The smart contract was deployed at Jan 7th, 2022. Only 11 days later, the trading volume and the number of transactions have reached up to \$944K and \$1.2K, respectively.
To gain a profit from players, \texttt{AniMoon} requires a certain amount of Ether when minting an NFT.
From Jan 8th to Jan 18th, 9,999 NFTs were minted, and the \texttt{AniMoon} team has earned 1,669 Ether, worth around \$6.3M at that time.
In addition, once an NFT is traded on \texttt{OpenSea}, the initiators of the project will receive a dividend, named \textit{creator earnings}~\cite{OpenSeaCreatorearnings}.
According to our collected transactions, the \texttt{AniMoon} team received approximately \$470K as the \textit{creator earnings} due to the trading of \texttt{AniMoon} NFTs on \texttt{OpenSea}.

\subsubsection{Running away and shutting down the project}
The rug pull occurs on Jan 18th, 2022, the 11th day since project launching.
Over \$6.3M worth of Ether was transferred out from the NFT smart contract.
All mint fees were withdrawn from the smart contract to developer controlled addresses. Interestingly, these addresses are hard-coded in the smart contract's \texttt{withdraw} function, which suggests that it is a premeditated scam.
Then, 1,560 Ether was transferred to \texttt{Binance} exchange and 109 Ether was transferred to \texttt{KuCoin} exchange.
In a nutshell, only after 20 days of launching, the volume of \texttt{AniMoon} has dramatically went down to \$66K, only around 7\% of its peak.
At May 15th, 2022, its \texttt{Twitter} account has been suspended, the official website was no longer being maintained, and the \texttt{Discord} server was also removed.
At last, all NFTs under the \texttt{AniMoon} collection become worthless tokens.
\section{Study Design}
\label{sec:data_collection}

 \begin{figure*}[t]
    \centering
    \includegraphics[width=0.95\linewidth]{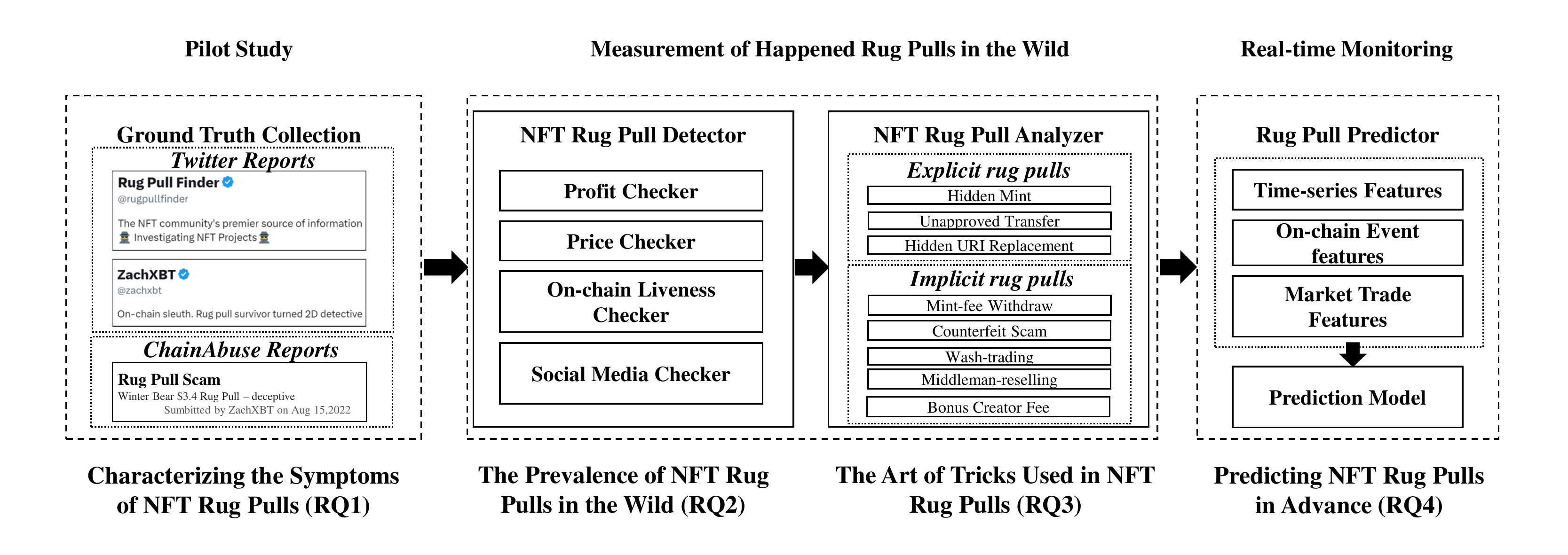}
    \vspace{-0.2in}
    \caption{The Overview of Our Study.}
    \vspace{-0.1in}
    \label{fig:overview}
\end{figure*}

In this work, we perform a series of progressive studies to gain a deep understanding NFT rug pulls. Specifically, we aim to answer the following research questions (RQs):

\begin{itemize}
    \item[RQ1] \textit{What are the characteristics of NFT rug pulls? Can we summarize concrete patterns to depict them?} Although NFT rug pulls are emerging in the ecosystem, no existing efforts have specified the concrete rules to distinguish whether an NFT project is a rug pull scam or not. Formulating the precise symptoms of NFT rug pulls is the key to identify them.
    \item[RQ2] \textit{How prevalent are NFT rug pulls in the wild?} Although media outlets have reported NFT rug pulls from time to time, there is a general lack of an understanding of \textit{to what extent NFT rug pulls exist in the wild}, and the impact of such scams.
    \item[RQ3] \textit{What are the tricks used in NFT rug pulls?} The success of an NFT rug pull depends heavily on the sophisticated tricks leveraged by the scammer. We, therefore, need to investigate the art of the tricks.
    \item[RQ4] \textit{Can we raise early warning of NFT rug pulls in advance?} Existing efforts usually rely on reactive methods to identify/flag NFT rug pulls after the scam has happened. However, an NFT rug pull usually involves several key steps in the whole process. Proactively identifying the rug pull projects in an early stage can help eliminate the potential risks they exposed.
\end{itemize}
%1) understand the characteristics of NFT rug pulls, 2) measure the prevalence of NFT rug pulls in the wild, 3) investigate the art of tricks leveraged in the NFT rug pulls, and 4) identify NFT rug pulls in their early stage and raise warning to the community in a proactive way.

Fig.~\ref{fig:overview} shows the overall process of our study. To understand the concrete symptoms of NFT rug pulls, we first harvest all the known NFT rug pulls revealed by our community, resulting a dataset of 253 NFT rug pull projects. 
Based on this dataset, we perform a pilot study to identify key symptoms of NFT rug pulls (\S~\ref{sec:symptom}). Taking advantage of the summarized symptoms, we design an effective detector to flag \textit{happened but undisclosed} rug pull scams in the wild by analyzing all the on-chain and off-chain data related to over 173K NFT projects (\S~\ref{sec:prevalence}). We have flagged 7,487 NFT rug pulls using the most reliable method, although it is only the lower-bound. Then, we investigate additional tricks used in these 7,487 scams, and design analyzers to automatically identify these tricks (\S~\ref{sec:tricks}). At last, to raise warnings of NFT rug pulls in advance (before they run away with investors' money), we devise a proactive method based on kinds of features for real-time monitoring, to raise warnings of the potential NFT rug pull projects in their early stages (\S~\ref{sec:predicting}).

\section{RQ1: The Symptoms of NFT Rug Pulls}
\label{sec:symptom}
\label{sec:rq1}
\subsection{Initial Ground Truth}
\label{sec:data:ground-truth}
Some entities in the community, e.g., security companies and researchers, are proactively engaged in reporting NFT rug pull scams. Their reports can be used as the initial source of our pilot study.
Specifically, we first harvest the labelled rug pull scams through \texttt{Chainabuse}~\cite{chainabuse}, a well known crowd-sourcing based threat intelligence platform, and two security research accounts on Twitter, i.e., \texttt{Rug Pull Finder}\footnote{\href{https://twitter.com/rugpullfinder}{https://twitter.com/rugpullfinder}} and \texttt{ZachXBT}\footnote{\href{https://twitter.com/zachxbt}{https://twitter.com/zachxbt}}, both of which constantly share reliable news of NFT scams.
After removing duplicates, we have collected 324 reports.
For further verification of these crowd-sourcing reports, we search the reported NFT projects in Google, intending to identify more evidences (e.g., scam accusation posts) supporting they are real scams. To the end, 253 NFT projects are labelled as rug pulls by at least two different sources, which are regarded as our initial ground truth.

\subsection{The Symptoms of NFT Rug Pulls}
\label{sec:rq1:characteristics}
%To precisely identify if there exists other rug pulled projects in the whole NFT ecosystem, characterizing exiting ones is the prerequisite. To this end, 
Based on the collected 253 rug pulls, we next summarize theirs common \textbf{s}ymptoms. To highlight their characteristics, we further select top-300 NFT projects from \texttt{OpenSea} top list~\cite{opensea_largest} for comparison. 
According to the general process of NFT rug pulls (see \S\ref{sec:background:rugpull}), we depict their symptoms from the following four aspects.

\begin{figure}[t]
\begin{tabular}{cc}
\begin{minipage}[t]{0.44\columnwidth}
    \includegraphics[width=1\textwidth]{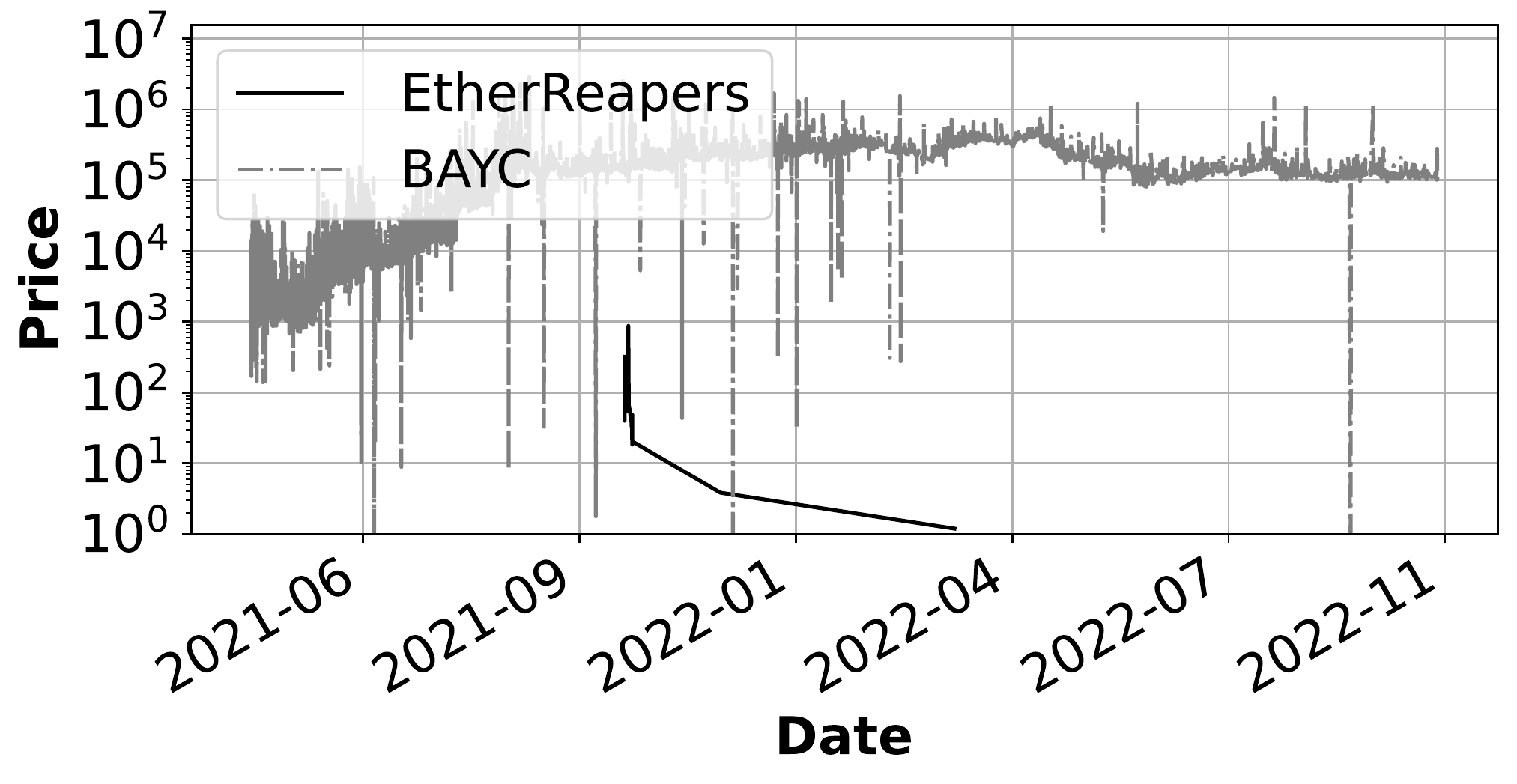}
    \vspace{-0.15in}
    \caption{The trading price of \texttt{EtherReapers} and \texttt{BAYC} in secondary markets.}
    \label{fig:pricecontrast}
\end{minipage}
\hspace{0.1in}
\begin{minipage}[t]{0.44\columnwidth}
    \includegraphics[width=1\textwidth]{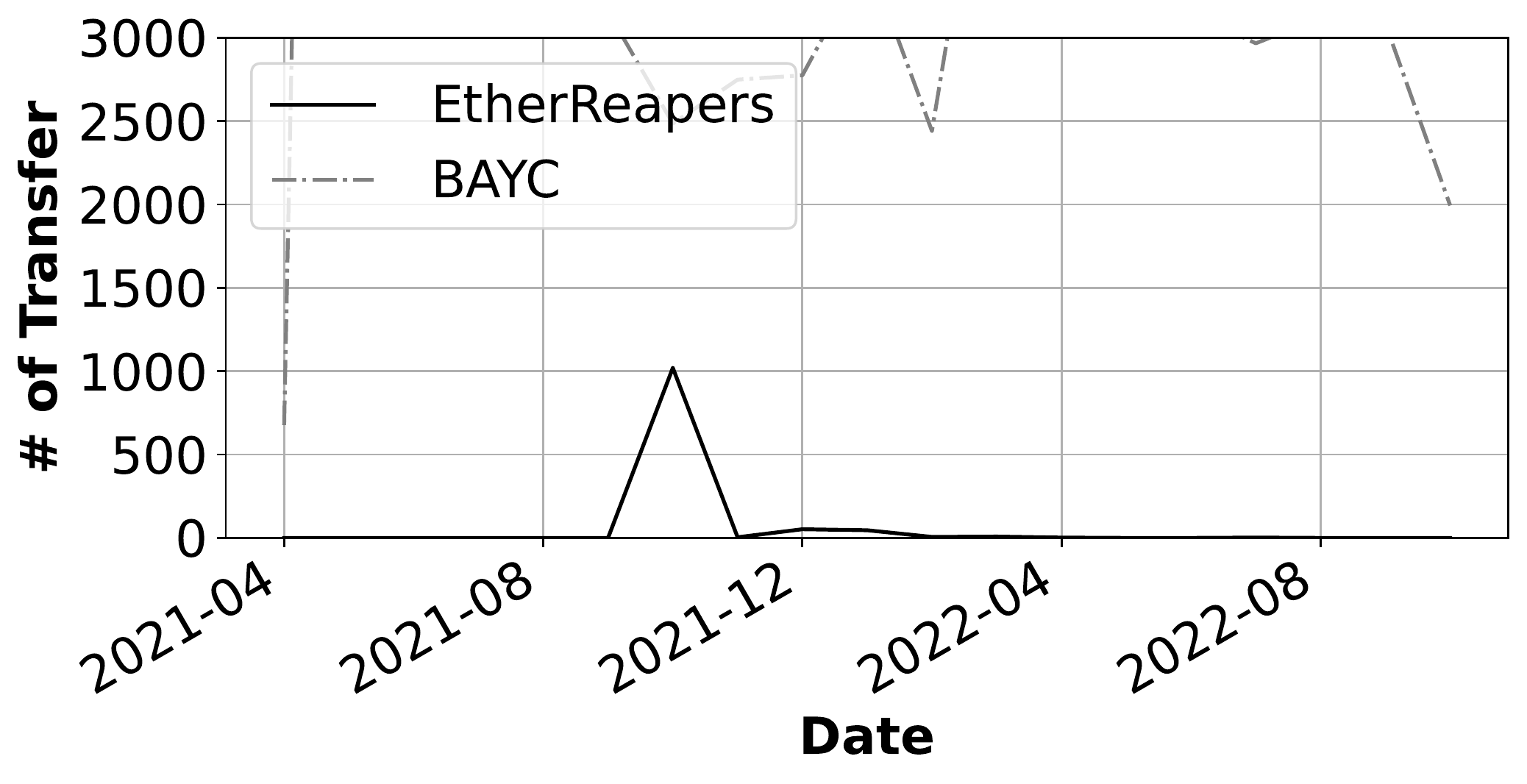}
    \vspace{-0.15in}
    \caption{The number of transfer events of \texttt{EtherReapers} and \texttt{BAYC}.}
    \label{fig:transfercontrast}
\end{minipage}
\end{tabular}
\vspace{-0.2in}
\end{figure}

\subsubsection{Profit Analysis}
Intuitively, making a profit is the underlying incentive of NFT rug pulls. Thus, a rug pulled project should be profitable. Our pilot study on the 253 confirmed NFT rug pull projects shows that all of them have made huge profits. In general, there are two ways of earning money for them. On the one hand, they can capture profit by requiring mint fee, and 251 of them belong to this category. Each of them has withdrawn 175 Ether from contracts on average.
The most aggressive one is \texttt{Apes In Space}\footnote{\href{https://etherscan.io/address/0x7a3b97a7400e44dadd929431a3640e4fc47daebd}{https://etherscan.io/address/0x7A3B97...}}, the rug puller has withdrawn around 2,645.82 Ether (roughly \$10.3M) through a single transaction.
On the other hand, 236 of them have been traded in secondary markets, which is also a way to make money. 
To be specific, the rug pullers can receive a sum of creator fee for each trade up to 10\% of the trading price.
Furthermore, the price in secondary markets could be easily pumped by the rug pullers. 
The owners could sell tokens with the high prices and gain profits by house wins.
To sum up, all the 253 projects have gained huge profits, which is a must-meet condition for a successful NFT rug pull.

\begin{itemize}
\item[\textbf{S1}] \textit{The creators of rug pull projects should gain a profit. This can be measured by either on-chain transactions or off-chain trade information in secondary markets.}
\end{itemize}

\subsubsection{Secondary Markets Price Analysis}
As discussed in \S\ref{sec:background:eth}, secondary markets play an important role in the life cycle of an NFT.
Fig.~\ref{fig:pricecontrast} shows the trading price of a rug pulled project \texttt{EtherReaper}\footnote{\href{https://etherscan.io/address/0x0a5b9b930fc5be638232d8b9b69cb5b46249c06e}{https://etherscan.io/address/0x0a5b9b...}} and a normal project \texttt{BAYC}\footnote{\href{https://etherscan.io/address/0xbc4ca0eda7647a8ab7c2061c2e118a18a936f13d}{https://etherscan.io/address/0xbc4ca0...}}.
Apparently, \texttt{EtherReaper} suffers an one-time dramatic rise and fall, while the price of \texttt{BAYC} is relatively stable although with some erratic fluctuation. Also, the price of \texttt{EtherReaper} will not be effectively recovered after the sharp drop. 
Furthermore, we can observe that \texttt{EtherReaper} has almost no trades after the dramatic decline in trading price, indicating that it has lost confidence from investors.

This symptom is not unusual. For those 236 ground truth projects that once have traded in secondary markets, 135 of them (57.2\%) remain ``silent'' after the rug pull, i.e., the NFT holders (victims) cannot sell any tokens after the scam. In addition, for the remaining 101 projects, over 90\% of them have extremely low prices currently (i.e., under 1\% of its top price), and the price does not rebound anymore.
As for the top-300 projects, even if the price may also fall (see Fig.~\ref{fig:pricecontrast}) with some erratic fluctuation, it will eventually rebound to normal in the end.
Such a huge gap between these two types of projects is highly associated with the floor price, which will be extremely low if liquidities are lost due to some reasons.
Intuitively, all top-300 projects still have active trades in the secondary market.
Thus, the evolution of trading activities can be regarded as an signal to flag rug pulled projects. We can conclude our second finding:

\begin{itemize}
\item[\textbf{S2}] \textit{Compared with non rug pulls, most of rug pull projects will eventually expire in secondary market after the scam, causing the extremely low liquidity and low token price, and the price could be never rebound.}
\end{itemize}

\subsubsection{On-chain Token Transfer Analysis}
Intuitively, the liveness of on-chain activities of rug pulled projects will decrease to an extremely low level after the rug pull.
Fig.~\ref{fig:transfercontrast} depicts the number of transfer events of \texttt{EtherReaper} and \texttt{BAYC}. 
For the rug pulled project, we can see a summit appears and then disappears after the rug pull scam happened. This follows the second stage mentioned in \S\ref{sec:background:rugpull}. However, this metric is quite stable for \texttt{BAYC}.
Therefore, we check token transfer events for rug pulled projects and top-300 projects.
For the former ones, the average number of on-chain transfer within the first month since their launching is 5,819.35. However, the number drops to 51.42 (0.8\% of the peak value) in the following month
As for top-300 projects, the number of on-chain transfer events is fluctuate within a reasonable range (usually 20\% to 80\% to its peak value).
Therefore, we can observe a huge gap on on-chain liveness between these two types of projects.
We can conclude our third finding:

\begin{itemize}
\item[\textbf{S3}] \textit{The number of on-chain token transfer will decrease to an extremely low level after the rug pull, which is a good indicator to identify the happened rug pulls.}
\end{itemize}

\subsubsection{Social Media Analysis}
We can observe from the case in \S\ref{sec:background:example} that a rug pull is highly associated with the status of its corresponding official social media.
To be specific, for 253 ground truth, 97\% of them have maintained social media accounts, e.g., \texttt{Twitter} accounts, \texttt{Discord} groups, \texttt{Instagram} pages, and official websites.
All of them have abandoned their social media accounts after the scam, e.g., suspending \texttt{Twitter} and taking down their official websites.
However, as for those normal projects, all of them have active social media events at the time of writing. 
Therefore, we can observe a huge gap, which can be used to distinguish rug pulls from innocent ones.
Hence, we can summarize the finding:

\begin{itemize}
\item[\textbf{S4}] \textit{The status of social media accounts of rug pulled projects will be abandoned, e.g., suspended or silent for a long period, after the scam.}
\end{itemize}

\begin{framed}
\noindent\textbf{Answer to RQ1}
\textit{
The NFT rug pulls share typical symptoms in aspects including profit capturing, price fluctuation, transfer evolution, and social media liveness. The symptoms offer opportunities for us to uncover more undisclosed rug pull scams in the wild.
}
\end{framed}

\section{RQ2: The Prevalence of Rug Pulls in the Wild}
\label{sec:prevalence}

Based on the summarized symptoms of NFT rug pulls, we further want to explore \textit{to what extent NFT rug pulls exist in the wild}. Thus, we first collect all the on-chain transactions and off-line data related to Ethereum NFT projects, and then we design an NFT rug pull detector to flag rug pull scams that have already taken place. Note that, we aim to get the most reliable result of NFT rug pulls, thus we have enforced a strict rule-based method in this section. Although the number of identified NFT rug pulls is a lower-bound, they are sufficient for us to measure the overall landscape (detailed in this section), summarize their tricks (\S\ref{sec:tricks}) and build the prediction model (\S\ref{sec:predicting}) in the following sections.

\subsection{Data Collection}
As aforementioned, NFT rug pull scams have typical symptoms which can be reflected by their on-chain transactions, secondary market trades and their social media activities. 
Thus, we make effort to create a comprehensive NFT dataset, which is listed in Table~\ref{tab:data}. 

\begin{table}[]
\centering
\caption{The statistics of our dataset (till Nov.1st, 2022).}
\label{tab:data}
\begin{tabular}{@{}rccc@{}}
\toprule
\multicolumn{1}{l}{\textbf{Items}}  & \textbf{ERC-721} & \textbf{ERC-1155} & \textbf{Total} \\ \midrule
\multicolumn{1}{l}{\textbf{Contract}}        & 145,865         & 27,508            & 173,373 \\
\multicolumn{1}{l}{\textbf{Transfer Event}}  & 157,930,255     & 17,808,803        & 175,739,058\\
\textit{mint}                                & 105,024,562     & 4,889,667         & 109,914,229 \\
\textit{burn}                                & 1,615,223       & 739,805           & 2,355,028 \\
\textit{swap}                                & 51,290,470      & 12,179,331        & 63,469,801 \\
\multicolumn{1}{l}{\textbf{Market Trade}\ \ \ \ \ \ \ \ \ \ \ }    & \multicolumn{2}{c}{30,235,520}                & - \\
\textit{OpenSea Seaport}                     & \multicolumn{2}{c}{8,227,449}        & - \\
\textit{OpenSea Wywern}                      & \multicolumn{2}{c}{20,126,896}       & - \\
\textit{LooksRare}                           & \multicolumn{2}{c}{334,660}          & - \\
\textit{X2Y2}                                & \multicolumn{2}{c}{1,546,515}        & - \\
\multicolumn{1}{l}{\textbf{Social Media}} & 19,110           & 3,396             & 22,506 \\
%\multicolumn{1}{l}{\textbf{Ground Truth}}     & 262              & 8                & 270 \\ 
\bottomrule
\end{tabular}
\end{table}

\subsubsection{Contracts \& On-chain Events}
\label{sec:data:transfer}
Collecting all related smart contracts and their transactions is necessary for our analysis.
We have obtained addresses of all contracts that involved in the latest 100K NFT transfer provided by \texttt{Etherscan}~\cite{etherscan}, a widely used Ethereum browser.
Among these addresses, we have identified 145,865 \texttt{ERC-721} smart contracts and 27,508 \texttt{ERC-1155} smart contracts, according to their unique function signatures.
For all these contracts, we also extract necessary metadata from the API provided by \texttt{ChainBase}~\cite{chainbaseAPI}, such as contract name, launch timestamp, and creator address.

As mentioned in \S\ref{sec:background:eth}, under both \texttt{ERC-721} and \texttt{ERC-1155} standards, invoking the function \texttt{transfer} will not only initiate a transaction, but also emit an event which will also be recorded on-chain.
Collecting these events can help us better understand the money flow.
Thus, we have deployed a client node \texttt{Geth}~\cite{geth}, and synchronized all blocks until Nov. 1st, 2022 from its very beginning (corresponding to the first 15,871,480 blocks).
Consequently, through these transactions, we have collected 157M and 17M pieces of \texttt{ERC-721} and \texttt{ERC-1155} token transfer events, respectively.
These transfer events can be further divided into three categories, i.e., \textit{mint}, \textit{burn}, and \textit{swap}.
Specifically, they can be distinguished by the value of specific fields. The \texttt{from} field of a mint event is the \textit{null} address, while the \textit{to} field is the dead address for burn events. The remaining ones can be categorized into swap events, indicating ownership transfer of NFTs.
% As the result, we have collected 151,704,877 \texttt{ERC-721} token transfer logs, 16,358,971 \texttt{ERC-1155} token transfer logs.

\subsubsection{Secondary Market Trades}
\label{sec:data:secondary}
%Trading plays a vital role in the NFT ecosystem. 
Different with \textit{token transfers}, referring to ownership transfer of NFTs between accounts, a \textit{trade} is only related to smart contracts owned by secondary markets and has nothing to do with the project contract.
To this end, we focus on all transactions interacted to smart contracts of the three dominant NFT markets, i.e., \texttt{OpenSea}~\cite{opensea_2022_opensea}, \texttt{X2Y2}~\cite{a2022_x2y2}, and \texttt{LooksRare}~\cite{_2022_looksrare}, accounting for 83.87\% of total history trading volumes in Ethereum~\cite{dappradarmarketplace} for \texttt{ERC-721} and \texttt{ERC-1155} NFTs.
We have conducted a comprehensive manual analysis on their deployed smart contracts and filtered out all related transactions to extract trading information.
Specifically, we firstly filtered out all external or public functions in these smart contracts to find ones that directly handle trading requests.
Then, we collected all events emitted by these functions.
Finally, we extracted all necessary information from these collected events, i.e., token id, token contract address, buyer's address, and seller's address.
Because trading NFTs often allows different cryptocurrencies, we use \texttt{Ethplorer}~\cite{a2022_ethplorer} to convert them to USD according to real-time interest rate.
Totally, we have collected around 30.2M pieces of trades.

\subsubsection{Social Media Activities}
\label{sec:data:social}
Our pilot study in \S\ref{sec:symptom} shows that as the first step of rug pulls, developers always promote their projects via social media to attract as many unsuspecting users as possible.
Moreover, the shut down of social media signals the end of the NFT rug pull to some extent.
We target \texttt{Twitter}, \texttt{Discord}, \texttt{Instagram}, and official NFT websites, to collect indicators of suspending or closing of such social media accounts.
Specifically, we get access to the API from \texttt{ChainBase}~\cite{chainbaseAPI}, to get related links about those social media platforms.
For \texttt{Twitter}, projects maintainer can make formal official announcements. 
To examine the status of the corresponding \texttt{Twitter} accounts, we queried the official APIs~\cite{TwitterAPI} based on the user names crawled from \texttt{ChainBase}.
In total, 19,110 \texttt{\texttt{ERC-721}} and 3,396 \texttt{ERC-1155} projects are linked to \texttt{Twitter}/\texttt{Discord}/\texttt{Instagram}, or their own official websites.

\subsection{Rug Pull Detector}
\label{sec:detectmethod}
Based on the typical symptoms we summarized in \S\ref{sec:rq1:characteristics}, we next measure the prevalence of NFT rug pulls in the wild.
To this end, we have designed and implemented a detector, which takes data of a project, e.g., historical transfer and trade transactions, and social media accounts status, as inputs, and outputs a report to show whether the examined project has been rug pulled or not.
The detector is composed of four independent components, each of which is responsible for examining a specific feature, e.g., price fluctuation and profitability.
A component will issue an alert if its corresponding rule is met.
If all these four independent components raise alarm simultaneously, the project will be labeled as being rug pulled.
Note that, our detector aims to identify the most reliable NFT rug pulls that have taken place using strict rules, which we believe is a lower-bound of the rug pull scam, based on which we can further train a rug pull predictor in \S\ref{sec:predicting}.

\subsubsection{Profit Checker}

As stated in \textbf{S1}, a success NFT rug pull project can grab profit in two ways, i.e., through on-chain Ether transfer or secondary market trades.
Therefore, we design our profit checker from these two aspects.
Specifically, we check on-chain logs to examine if there exist transactions sent Ether to the project. This indicates the contract creator behind the project receives profits directly.
Otherwise, we firstly check whether a project can gather profits from secondary markets by analyzing the rewarding mechanism of each market. For example, if an NFT is traded successfully on \texttt{OpenSea}, the project will receive a sum of \textit{creator fee} that can be withdrawn later.
If the project can meet the requirement for one of the two above, then we assume this projects is profitable.
For the sake of covering all possible profit-making activities, in the profit checker, we only focus on whether the project has profitable behavior, without considering how much profit it actually makes.
In addition, even if the Ether is not directly transferred to the contract, or the owner of the contract, we also assume it is profitable, as there exist projects gain profit by middleman (see \S\ref{sec:middleman}). This checker could help identify all the rug pull candidates for a further verification by other three checkers we implemented.

\begin{figure*}[t]\centering 
    \subfigure[The number of daily transactions.]{ 
\begin{minipage}{0.30\textwidth}\centering
\includegraphics[width = \textwidth]{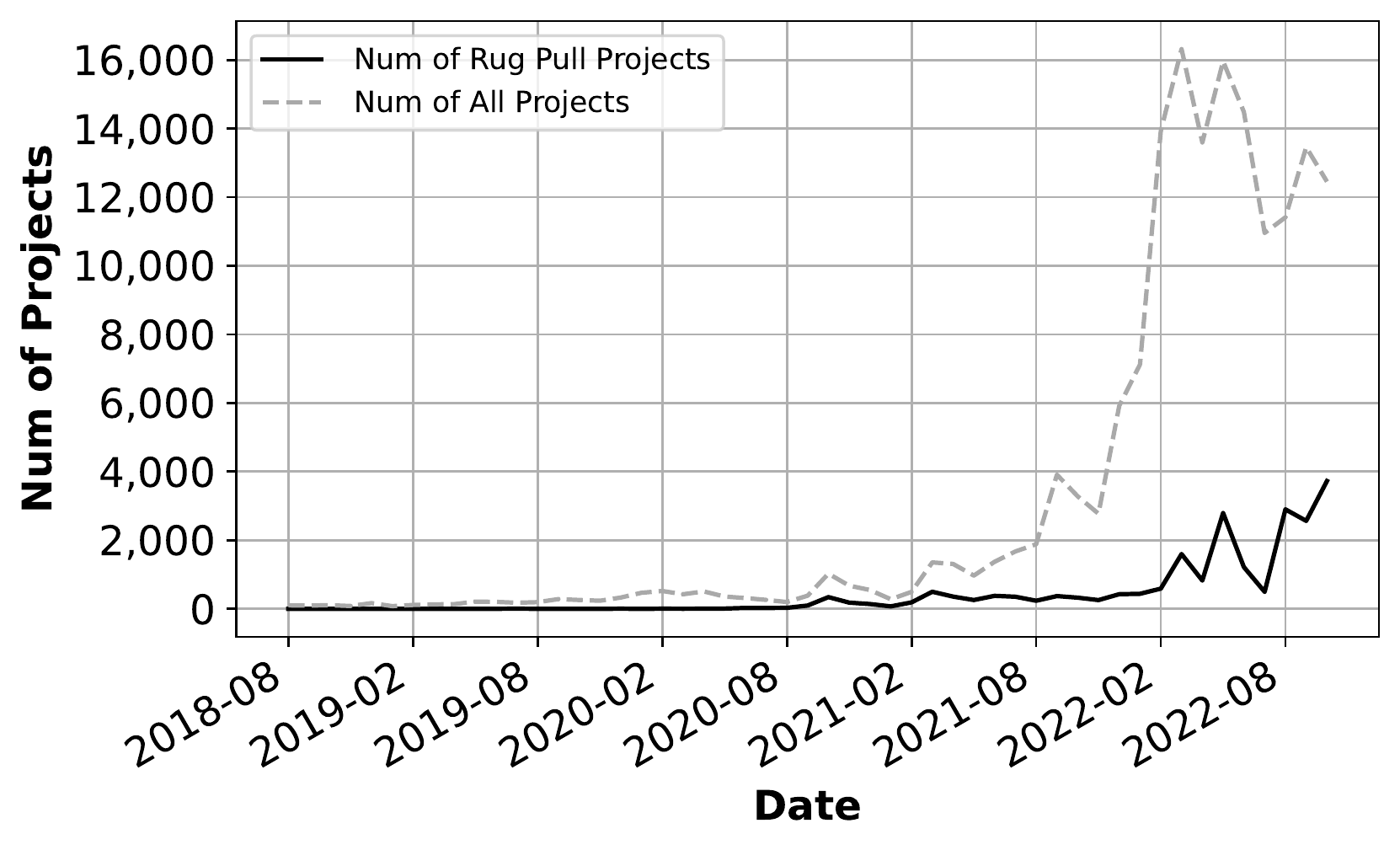} 
\end{minipage}}
\subfigure[The number of daily events.]{ 
\begin{minipage}{0.30\textwidth}
\centering 
\includegraphics[width = \textwidth]{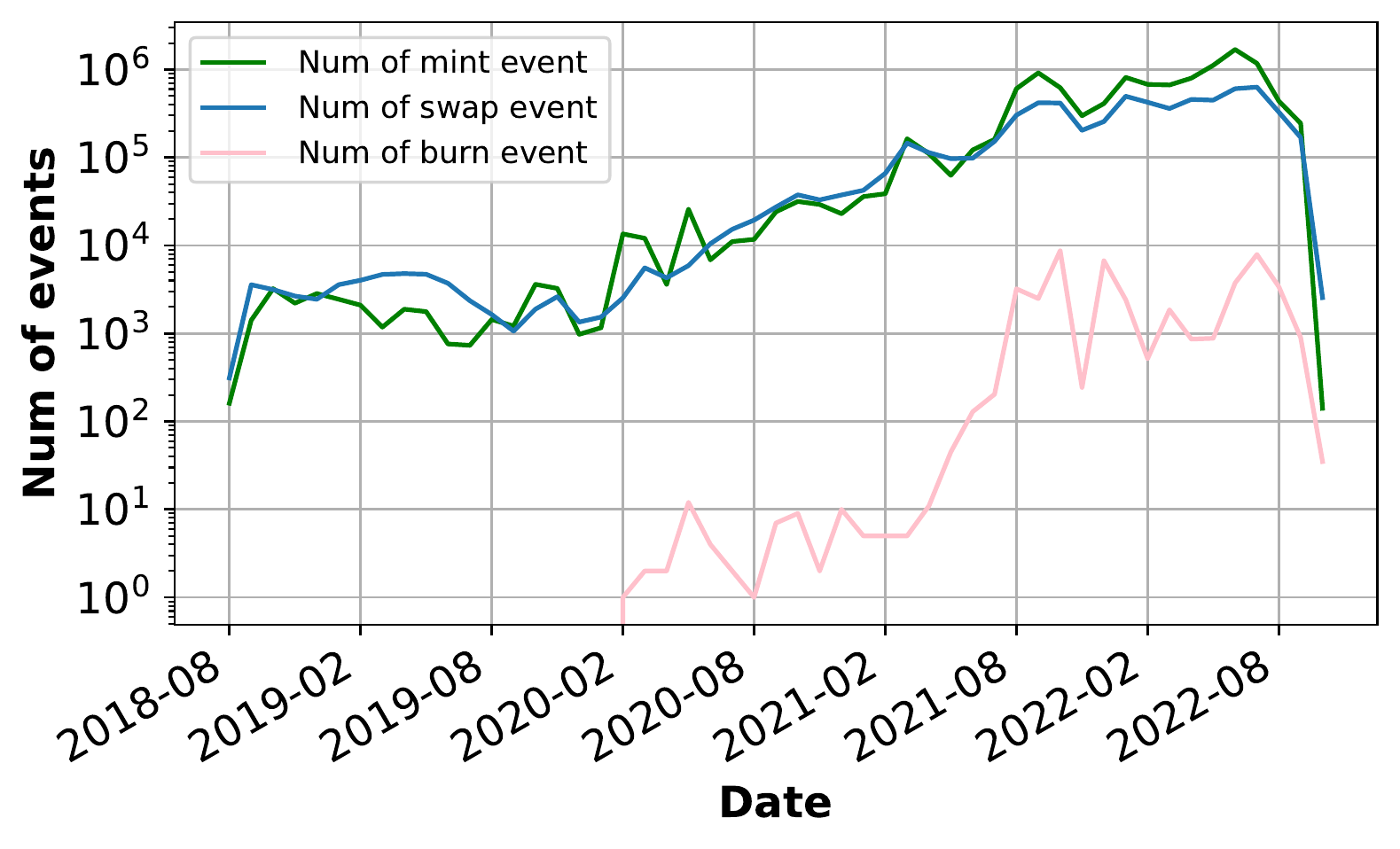} 
\end{minipage}
}
\subfigure[The number and trading volumes of daily trades.]{ 
\begin{minipage}{0.30\textwidth}
\centering 
\includegraphics[width=\textwidth]{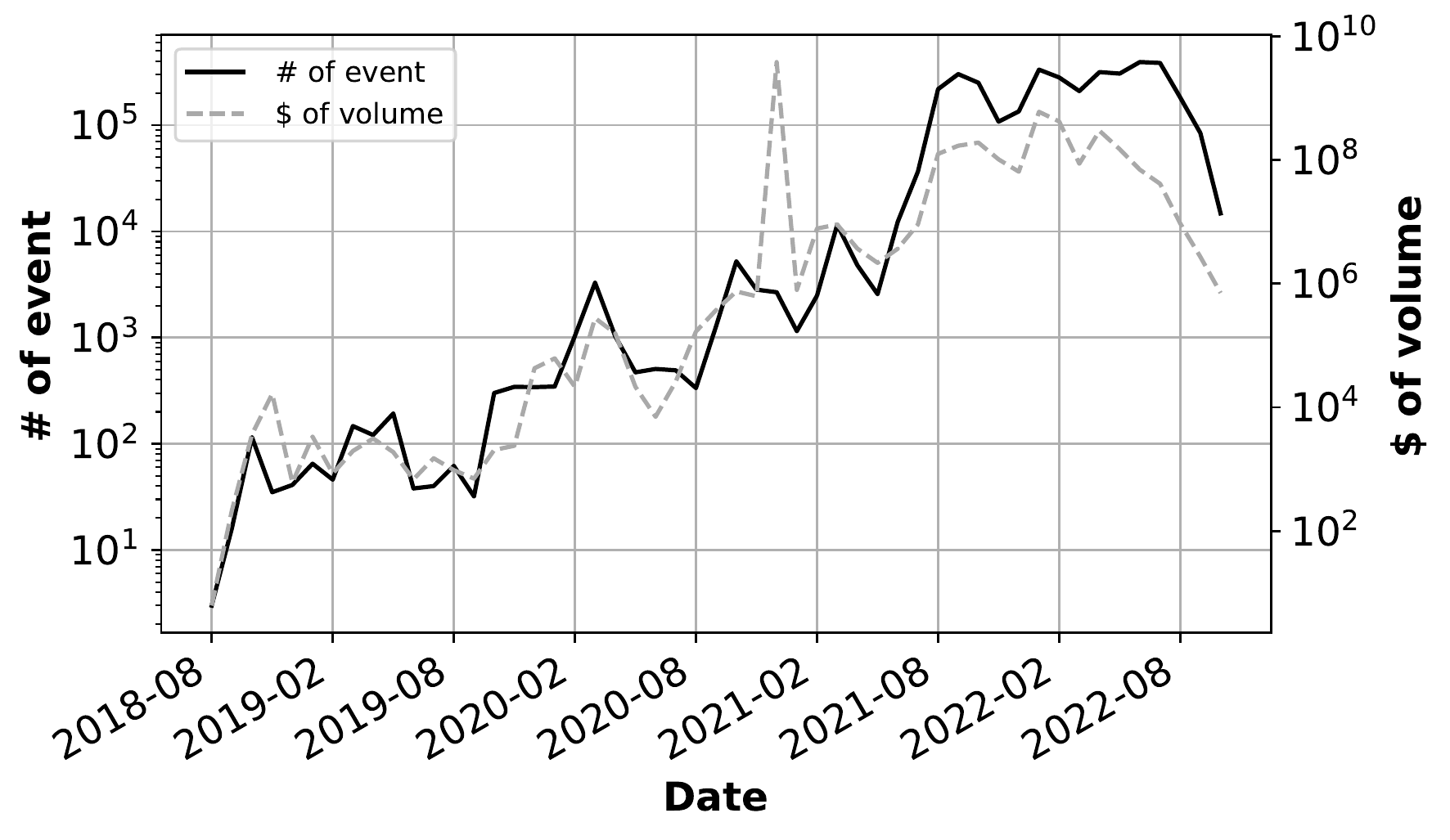} 
\end{minipage}
}
    \vspace{-0.1in}
\caption{General overview of rug pulled projects.}
    \vspace{-0.1in}
\label{fig:general_overview}
\end{figure*}

\subsubsection{Price Checker}
According to the solid line in Fig.~\ref{fig:pricecontrast} and \textbf{S2} in \S\ref{sec:rq1}, the trading prices of rug pulled projects in secondary markets have two obvious characteristics. On the one hand, the price will be pumped to the summit, and dramatically declined due to a rug pull occurs. On the other hand, the price of NFTs of rug pulled projects will not rebound.
Therefore, we introduce two metrics to quantify these two characteristics, i.e., \textit{drawdown} and \textit{recovery}, respectively.

Specifically, for each project, according to its trading history in secondary markets, we construct a chronological trading price sequence $\mathcal{P}$. $\mathcal{P}$ consists of a series of $p_{time}^{token}$, where $time$ and $token$ refer to the timestamp and the token identifier, respectively, and $p$ is the trading price in USD. For example, $10_{1640995200}^{3}$ refers to the NFT indexed by $3$ traded by \$10 at the midnight of Jan. 1st, 2022.
Note that, for any two adjacent items, the later one strictly happens after the former one, though they may correspond to different NFTs under the same project.
For each $p_{i}$ in $\mathcal{P}$, its $drawdown$ is defined as:
\begin{equation}
drawdown_i = max(\frac{p_i-p_j}{p_i}), \text{where}\ j > i \nonumber
\end{equation}
where $p_{j}$ corresponds a trade happens strictly after the given $p_{i}$. In other words, the $drawdown_{i}$ is calculated by $p_{i}$ and the \textit{lowest trading price} after it.
To this end, we can construct a \textit{drawdown sequence}, whose length equals to $|\mathcal{P}|-1$.
Heuristically, if any element in the drawdown sequence is greater than 0.99, we can conclude that there is a dramatic price plummeting. For example, for \texttt{EtherReapers} depicted in Fig.~\ref{fig:pricecontrast}, the highest trading price happened on it first trade, which corresponds a $drawdown$ around 0.9963.
%because the lowest trading price is only 1.19 USD occurred in the last trade.

Moreover, we can also observe a long tail in Fig.~\ref{fig:pricecontrast}, which reflects the unrecoverability of all NFTs of a rug pulled project as mentioned in \textbf{S2}.
Thus, for $drawdown_{i}$ whose value is greater than 0.99, we further examine if its $p_{j}$ can be traded with a much higher price.
Let us assume the $p_{j}$ is related to the token $t$, i.e., $p_{j}^{t}$. Its $recovery_j$ can be defined as:
\begin{equation}
recovery_j = max(\frac{p_k^{t}-p_j^{t}}{p_j^{t}}), \text{where}\ k > j \nonumber
\end{equation}
, where $p_{k}^{t}$ is the price that corresponds to the following trades on the same NFT. 
Similarly, we heuristically set the threshold of $recovery_j$ as 0.01, indicating the highest trading price after a local minimum $p_j$ can only be 1\% higher for the same token.
Take \texttt{EtherReapers} as an instance again, we have identified 60 $drawdown$ greater than 0.99 in total. For all $p_{j}^{t}$ in these $drawdown$, the highest one is 0.9986, related to the NFT whose token ID is \texttt{0x01b9}\footnote{\href{https://etherscan.io/tx/0x1bfc75769c9f4b2035537fce9e4c22cc574a5a751b03bcff393ecc394d42e950}{https://etherscan.io/tx/0x1bfc75...}}.
 However, as for all related tokens that result in the $drawdown$ over 0.99, there are no other trades of these tokens. 
In other words, after a sharp drop in prices, the price of all NFTs cannot be recovered at all.
Note that, if $t$ is not traded at all after $p_{j}^{t}$, we set the $recovery_j$ as 0 due to the lack of liquidity.

\subsubsection{On-chain Liveness Checker}
The liveness of an NFT project, summarized by \textbf{S3}, can be reflected by the number of on-chain transactions.
Specifically, we denote $N_1$ and $N_2$ as the number of transfer happened in the first month once the first token was minted and the most recent month, respectively.
Based on them, we define the \textit{on-chain liveness} as follows:
\begin{equation}
\label{equ:liveness}
liveness = \frac{N_1-N_2}{N_1}\nonumber
\end{equation}
, where the $liveness$ should be positive and greater than 0.99, meaning that its liveness is neglectable compared to the first month in October, 2022, which may due to suffering rug pulls.
For example, the transfer related to \texttt{EtherReaper} in October is only 0.8\% of its first month.

\subsubsection{Social Media Checker}
Shutting down the social media services is an effective evidence of NFT rug pulls. 
As mentioned in \S\ref{sec:data:social}, we have tried our best to collect all the social media accounts related to the NFT projects through APIs provided by \texttt{ChainBase}~\cite{chainbaseAPI}.
We next check the status of their social media accounts.
For \texttt{Twitter} account, we check if it is suspended, deleted, or inactive for a long time (at least one month). 
Furthermore, we check whether the invitation link is expired for \texttt{Discord}, whether the account can still be found or be suspended for \texttt{Instagram}, and whether the server is down for their official websites.
If any of the above requirement is satisfied for a specific project, we consider it as a possible rug pull.
Note that, for those projects do not have any social media, we do not label them as rug pulls, as our pilot study suggests that advertising plays an important role in the life cycle of rug pulls. We admit that this may miss several rug pull scams, while \textit{our detector in this section only wants to expand the ground truth of rug pulls in a most reliable way}.

\noindent
\textbf{Accuracy of the Detector.}
Before analyzing all the NFT data, we first evaluate the accuracy of our detector. We admit that rule-based detection shares inherent limitations, e.g., rely heavily on existing knowledge. 
We feed all the 253 ground truth NFT scam projects and top-1000 popular NFT projects on \texttt{OpenSea} to our detector, and observe that 232 of 253 rug pull projects can be identified correctly, i.e., \textit{no false positives} and 21 false negatives remain.
This is because 10 of them lack social media information, so they cannot be alerted by our social media checker.
For the remaining 11 false negatives, we found that they were not raised alarms by the price checker. Specifically, the \textit{drawdown rates} of them are roughly 90\% but not 99\% used in our checker.
However, as aforementioned, we aim to identify the most reliable NFT rug pulls (i.e., flag no false positives), thus we have enforced most strict rules. 

\subsection{The Prevalence of NFT Rug Pulls}
\label{sec:overall-results}
We next apply our detector to all 145,865 \texttt{ERC-721} and 27,508 \texttt{ERC-1155} projects, aiming to characterize the overall landscape of NFT rug pull scams.

\subsubsection{Detection results}
\label{sec:overall-results:results}
Our rug pull detector labeled 7,487 NFT rug pull projects in total, with 7,019 \texttt{ERC-721} projects and 468 \texttt{ERC-1155} projects, accounting for 5.08\% and 1.70\% of total existing projects, respectively. We further sampled 50 \texttt{ERC-721} projects and 50 \texttt{ERC-1155} projects for manual verification, we did not observe any false positives. We also disclose our detection result to an anonymous leading blockchain security company, and they have confirmed our findings. It suggests the reliability of our approach, although it only reports the lower-bound.

Among them, 5,601 \texttt{ERC-721} contracts and 84 \texttt{ERC-1155} contracts require a sum of mint fee to mint tokens, a way of grabbing profit.
According to our statistics, they totally mint around 11.3M \texttt{ERC-721} tokens and 93K kinds of \texttt{ERC-1155} tokens. 
In addition, these labeled projects correspond to 17.9M pieces of transfer events recorded on-chain, and 3.6M pieces of trade events on secondary markets, reaching up to \$6.2B in terms of trading volume. Although the number is impressing, we will show that the scammers tend to create the mirage of a prosperity project by manipulating the trade events in \S\ref{sec:tricks}.
They are created by 6,778 addresses, which can be regarded as the real culprit that conducts rug pulls.
Most rug pullers (92.36\%) create only one project, while the other 518 ones operate multiple rug pull projects.
Astonishingly, the address \texttt{0x453e}\footnote{\href{https://etherscan.io/address/0x453e23826f0cff7655b6a7e866123013923ae818}{https://etherscan.io/address/0x453e23...}} have created 12 projects, all of which have been rug pulled. 

\subsubsection{Trending of NFT rug pulls}
\label{sec:trending}
Fig.~\ref{fig:general_overview} shows an overview of these 7,487 happened rug pulls.
As shown in Fig.~\ref{fig:general_overview}(a), the trend of launching of rug pulled projects is consistent with the rising of NFT ecosystem, which becomes popular and grows rapidly at the beginning of 2022.
At the peak, there are over 50 projects that are created within a single day but finally being rug pulled.
Fig.~\ref{fig:general_overview}(b) shows the number of emitted events of rug pulled projects per month. As we can see, both the \textit{mint} and \textit{swap} events are constantly upward till August, 2022, which indicates that rug pulls have become more and more rampant. In 2022, on average, there are over $10^4$ pieces of mint events and swap events, respectively, emitted by these projects per day. Interestingly, the number of mint events is higher than the one of swap events, which means that each NFT token will only be swapped less than once on average. This finding is consistent with the low liquidity of rug pulled projects in the NFT ecosystem.
Moreover, as we mentioned in \S\ref{sec:data:transfer}, a \textit{burn} event indicates that an NFT is intentionally destroyed by its owner. Its number is also went upward from 2021.
However, we can observe a sudden decrease for all these three events after August.
This is because we only present the data until the end of October, while some projects are rug pulled in the following November.
Fig.~\ref{fig:general_overview}(c) shows the statistics of daily trades in terms of number and volumes.
In 2020,  the number of daily trade is 53.32 on average, while in 2022, it is 8,217.91. As can be seen, rug pulls have become more rampant in secondary markets than it was in 2020.

\begin{framed}
\noindent\textbf{Answer to RQ2}
\textit{Our detector has flagged 7,487 rug pulled NFT projects in the ecosystem, 29.6$\times$ greater than our collected initial ground truth. NFT rug pulls are becoming more and more prevalent in the ecosystem.
}
\end{framed}
\section{RQ3: The Art of NFT Rug Pulls}
\label{sec:tricks}

Beyond the pump and dump nature of NFT rug pulls, we further observe some tricks leveraged by scammers for facilitating the overall scam process. Specifically, we went though the scam reports of all the 253 initial ground truth, and further sampled 200 new NFT rug pulls identified by our detector, to manually investigate their transactions and smart contracts, aiming to uncover kinds of tricks used by them. 
These tricks can be classified into two categories, i.e., \textit{explicit tricks} (i.e., take advantage of backdoors in smart contracts to gain profit) and \textit{implicit tricks} (indirectly gain profits by performing price manipulation on the market), with eight kinds of tricks in total. 
For these kinds of tricks, we further implement automatic analyzers to explore their prevalence among the 7,487 flagged NFT rug pulls in the previous step.

\subsection{Explicit Tricks in NFT Rug Pulls}
\label{sec:rq2:explicit}
We have summarized three general patterns of smart contract backdoors favored by NFT scammers.

\begin{lstlisting}[caption={An example of a hidden mint backdoor. },label={fig:overmint},language=Java]      
totalSupply = 8390
maxTokens = 50000

function mint(address recipient, uint256 seed) external override whenNotPaused {
    require(admins[_msgSender()], "Only admins can call this");
    require(minted + 1 <= maxTokens, "All tokens minted");
    minted++;
    generate(recipient, minted, seed);
    ...      
    _safeMint(recipient, minted);
}
\end{lstlisting}

\subsubsection{Hidden mint}
Hidden mint refers to the behavior where NFT projects privately mint tokens whose number is beyond the maximum predefined number.
As we mentioned in \S\ref{sec:background:eth}, NFT value is totally determined by the relationship of demand and supply.
Thus, such a hidden mint is only beneficial to gain extra profits for rug pullers, and extremely harmful to holders.
Typically, there is a field, named \texttt{totalSupply}, which will return the maximum allowable minted number of tokens.
However, some rug pullers break this limit by exploiting backdoors hidden in the smart contract to overmint tokens and earn extra profits.

\textbf{Case Study.} 
Take a famous NFT rug pull project, \texttt{Cat\&Mouse}\footnote{\href{https://etherscan.io/address/0xe19e0cb95a9e39cb0ecde82e8a9f9f432835a0d0}{https://etherscan.io/address/0xe19e0c...}}, as an example, whose code snippet is shown in Listing~\ref{fig:overmint}.
As we can see, \texttt{totalSupply} at L1 declares that at most 8,390 NFTs can be minted and circulated. However, we have observed 8,879 available NFTs circulated on-chain, which is resulted from the implementation of \texttt{mint} (L4).
Specifically, L5 asserts that only the designated administrators have the mint permission. At L6, it examines whether a mint action is allowed by comparing to the upper limitation, which is declared by \texttt{maxTokens} (set as 50,000) instead of \texttt{totalSupply}.
It means that only the designated administrators are allowed to overmint NFTs, which will extremely devalue minted NFTs.

\textbf{Detection.}
To pinpoint how many NFT rug pulls have exploited hidden mint, we firstly fetch the declared value of \texttt{totalSupply} of all NFT contracts from APIs provided by \texttt{Alchemy}~\cite{alchemyapi}.
Then, we trace all the transactions initiated from each project, and calculate whether the amount of NFTs in circulation is greater than the value of \texttt{totalSupply}.
Because the non-fungibility of \texttt{ERC-721} tokens, simply calculating circulated amount is feasible.
As for NFTs following \texttt{ERC-1155}, however, because a token ID can correspond to multiple identical tokens, the same method is not applicable.
As a result, we found 628 \texttt{ERC-721} projects (8.4\% of all NFT rug pulls we labelled) have minted tokens covertly.

\begin{lstlisting}[caption={An example of an unapproved transfer.},label={fig:unaothorzied},language=Java]  
function setContracts(address _staking, address _metadata) external onlyOwner {
    staking = IBlockverseStaking(_staking);
    metadata = IBlockverseMetadata(_metadata);
}    

function transferFrom(address from, address to, uint256 tokenId) public virtual override(ERC721, IERC721) {
    // allow admin contracts to be send without approval
    if(_msgSender() != address(staking) && _msgSender() != owner()) {
        require(_isApprovedOrOwner(_msgSender(), tokenId), "ERC721: transfer caller is not owner nor approved");
    }
    _transfer(from, to, tokenId);
}
\end{lstlisting}

\noindent
\subsubsection{Unapproved transfer}

An unapproved transfer indicates that developers can transfer NFTs on behalf of someone arbitrarily without his authorization.
In various standards, including \texttt{ERC-721} and \texttt{ERC-1155}, transferring tokens to someone on one's behalf is normal, however, it should be properly authorized. There are two authorization-related functions should be implemented in \texttt{ERC-721} and \texttt{ERC-1155} contracts, i.e., \texttt{approve} and \texttt{setApprovalForAll}, which allow a user to approve someone to transfer NFTs on behalf of him.
If any of these two functions is implemented incorrectly accidentally, or even deliberately, attackers may arbitrary transfer others NFTs out for different goals, e.g., gaining extra profits.

\textbf{Case Study.}
Listing~\ref{fig:unaothorzied} shows a concrete example from \texttt{Blockverse}\footnote{\href{https://etherscan.io/address/0xb9d9455ea8ba8e244b3ea9d46ba106642cb99b97}{https://etherscan.io/address/0xb9d945..}}, which has completed 1,310 pieces of unapproved transfers.
As we can see, the variable \texttt{stacking} at L2 can be arbitrarily assigned by the owner. In \texttt{transferFrom}, it first identifies if the transaction initiator is the address declared by \texttt{stacking} (L8). If it is, the authorization step at L9 will be passed.
To this end, unauthorized transfers can be performed under the acquiescence of the owner.

\textbf{Detection.}
To identify such behaviors, we utilize all collected transfer events.
Specifically, invoking \texttt{approve} and \texttt{setApprovalForAll} will also emit events, which consist of the initiator, the authorized user, and the token ID(s).
Therefore, we collect all the approve events of rug pulls and construct a \textit{directed graph}, whose nodes are users, edges are directed from the initiator to the authorized one. 
Each edge corresponds to a list, composed of approved token ID(s) and the corresponding timestamp.
Note that, each approval can be withdrawn by its initiator.
To this end, for each successful token transfer event, e.g., \texttt{Bob} initiates a \texttt{transferFrom} where \texttt{from} is \texttt{Alice} and \texttt{to} is \texttt{Carol}, we traverse the the directed graph to find if there is an edge directing from \texttt{Alice} to \texttt{Bob}, and the edge contains the transferred token ID where the approval happened before the \texttt{transferFrom}.
In total, 
we found 1,335 pieces of unapproved transfer, related to 9 NFT projects.

\begin{lstlisting}[caption={An example of hidden URI replacement.},label={fig:setTokenURI},language=Java]      
function setTokenURIExtension(uint256 tokenId, string calldata uri) external override extensionRequired {
    _setTokenURIExtension(tokenId, uri);
}

function setTokenURIExtension(uint256[] memory tokenIds, string[] calldata uris) external override extensionRequired {
    require(tokenIds.length == uris.length, "Invalid input");
    for (uint i = 0; i < tokenIds.length; i++) {
        _setTokenURIExtension(tokenIds[i], uris[i]);            
    }
}

function _setTokenURIExtension(uint256 tokenId, string calldata uri) internal {
    require(_tokensExtension[tokenId] == msg.sender, "Invalid token");
    _tokenURIs[tokenId] = uri;
}
\end{lstlisting}

\noindent
\subsubsection{Hidden URI replacement}

The term URI stands for \textit{universal resource identifier}, through which the bound real-world items of NFTs can be accessed.
For an NFT, its whole value is determined by the item linked to its URI. In other words, if the URI is stealthily replaced, the NFT's holder may suffer huge financial loss.

\textbf{Case Study.}
Listing~\ref{fig:setTokenURI} shows  \texttt{Cryptobiotica}\footnote{\href{https://etherscan.io/address/0x28b99de53a32bf715353886090dab8a8652fe13b}{{https://etherscan.io/address/0x28b99d...}}}, an example that can replace the URIs.
Codes from L12 to L15 is the function \texttt{\_setTokenURIExtension} that can replace the corresponding URI of the given NFT. It can be called by two functions declared at L1 and L5, where the latter one can perform a batch replacement.
According to statistics, we observe that each of these two functions has been called twice, and 155 URIs are replaced by the owner stealthily.

\textbf{Detection.}
To determine whether a URI replacement is likely to happen, we apply \texttt{Slither}~\cite{slither}, a widely used static analysis tool for Ethereum smart contracts, on NFT contracts and trace all global variables related to URI.
Then we traverse all the usage of these global variables and check whether they are modified (e.g., privileged backdoors). 
If it does, then we assume the contract is vulnerable to hidden URI replacement.
As a result, we find that 997 NFT projects (13.3\%) have a privileged function, named \texttt{setTokenURI}, which can only be called by the owner of the contract. After tracing all transactions interacted with these contracts, we discovered that 30 of them have been called to replace URI.
The most aggressive NFT project,  \texttt{scorpio.world}\footnote{\href{https://etherscan.io/address/0x2e2691819d3441994e9709e776bd77d08cd9e89b}{https://etherscan.io/address/0x2e2691...}}, has been called 781 times of URI replacement, which is the largest.

\subsection{Implicit Tricks in NFT Rug Pulls}
\label{sec:rq2:implicit}
Apart from explicit rug pulls, which could be identified by code auditing, implicit rug pulls refer to the behaviors that scammers manipulate the NFT market for fun and profit. 
We have summarised five such tricks as follows.

\subsubsection{Mint fee withdraw}

Withdrawing mint fee refers to a situation that the contract requires users to mint NFTs by charging mint fee, which can be withdrawn by the owner as profits.
However, it is possible for a founder to withdraw all mint fee to perform a rug pull.
If this happened on NFT projects, both secondary markets and holders will lose confidence of them.

\textbf{Detection.}
Minting a token will generate a transfer event whose \texttt{from} is a null address. Taking advantage of this characteristic, we extract such transfer events whose \texttt{from} is a null address and \texttt{value} is not zero. Then, we will observe if the withdraw function is called, where the owner transfers all balance out. In total, we have found 4,821 contracts (64\%) that have withdrawn mint fee. 
Table~\ref{tab:mintfee-withdraw} in Appendix shows the top 5 NFT projects in terms of withdrawn Ether.
As we can see, rug pullers can gain huge profits (over thousands of Ethers) for each project.
Interestingly, the first-most profitable project, named \texttt{Apes In Space}, has been withdrawn only once, while the rug puller obtained over \$10M through a single transaction.

\subsubsection{Counterfeit NFT}
\label{sec:conterfeit:scam}
Because the value of an NFT collection is highly dependent to its reputation, some malicious developers try to counterfeit a fake one that is confusingly similar to a popular NFT project.
Not only do they name it with a similar or even identical name, but they also bind NFTs with real-world items in similar style, e.g., similar painting styles and objects in figures, to mislead unsuspecting investors.

\textbf{Detection.}
To effectively identify such scams, we apply the \textit{Levenshtein distance}~\cite{levenshteindistance} to measure the similarity of two NFT project names.
A high Levenshtein distance between two names suggests the names are highly similar.
For normalization, we introduce the the \textit{Levenshtein ratio} as follows:
\begin{equation}
    ratio_{lev}=\frac{(len(a)+len(b))-lev(a,b)}{len(a)+len(b)}\nonumber
\end{equation}
, where $len(a)$ and $len(b)$ is the string length of $a$ and $b$, and $lev(a,b)$ is the Levenshtein distance between $a$ and $b$. Consequently, the Levenshtein ratio will be 1 if two candidates are identical.
We compute Levenshtein ratio between collected 7,487 NFT projects and the top-1000 NFT projects on \texttt{OpenSea}~\cite{OpenSeatop}.
Compared to those famous projects, there are 14 rug pulled projects with $ratio_{lev}$ over 95\%, and 68 projects with $ratio_{lev}$ over 90\%.
Moreover, there are 6 projects even adopt the exact same name with the popular ones.
For example, two projects are named as \texttt{MUSHROHMS}, while the popular one\footnote{\href{https://etherscan.io/address/0x133ba8f869f3ae35a5ca840ba20acfa31b0e2a61}{https://etherscan.io/address/0x133ba8...}} has 1,546 market trades for over \$5M since its creation, and the fake one\footnote{\href{https://etherscan.io/address/0x3760513ef525df09169497b5b79176e1a830a773}{https://etherscan.io/address/0x376051...}} has no market trade at all.
Interestingly, around 94.75\% identified rug pulled projects have low Levenshtein ratio (under 0.8) compared to those popular ones. We can conclude that most rug pullers still intend to create their own brands.

\subsubsection{Wash trading}
Wash trading refers to the behavior that NFTs are traded within a group of holders to make a fake prosperity for the corresponding project.
Wash trading can be used in a type of notorious attack, i.e., \textit{pump \& dump}~\cite{xu2019anatomy}.  
Specifically, victims may be trapped due to its prosperity, suffering financial loss after a rug pull. 
\textbf{Detection.}
To identify wash trading behaviors in the flagged rug pull scams, we take advantage of its key feature, i.e., tokens are usually traded between two users repeatedly.
Thus, for each NFT project, we built a direct graph, where nodes are composed of participants of the project. Once user \texttt{Alice} sells an NFT to user \texttt{Bob}, we add an edge directed from \texttt{Alice} to \texttt{Bob}, where the amount paid in USD is linked to the edge.
If there are more than 10 edges between any two accounts within a project, we heuristically label the \textit{seller-buyer} pair is used for wash trading.
As a result, we found 26 rug pulled projects performed 2.7K times of wash trading, accounting for more than \$1.04B in terms of history trading volumes in secondary markets.
The top 5 (shown in Table~\ref{tab:washtrading} in Appendix) out of those 26 projects accounts for 99.04\% of the total in terms of trading volumes.
As we can see, the most influential one has performed wash trading up to 735 times, corresponding to more than \$377M.
Such a huge amount of trading volumes will inevitably mislead investors.

\subsubsection{Middleman reselling}
\label{sec:middleman}

Middleman reselling means that all the NFTs of a project are minted to a single account, who then resells them on secondary markets. 
For potential buyers, they may think that the person they are dealing with is a whale account who holds bunch of NFTs.
However, the seemingly \textit{whale account} has high likelihood to be a collusion address controlled by the rug pullers.

\textbf{Detection.}
To identify such a behavior, we extract all mint events and calculate the distribution of minted NFTs.
As a result, we find that 865 rug pulled NFT projects (11.6\%) mint all tokens to a single account.
Among these, NFTs of 273 projects have not been involved in any on-chain transactions at all. As for the remaining 592 projects, NFTs belonging to these projects have not been traded in any secondary markets. However, this does not mean that these projects are not profitable, and we presume that they did this deliberately.
The reason we inferred this is that we found 70 of them have direct Ether transfer events from some user addresses.
Thus, we regard these 70 rug pull projects exploited the middleman reselling.
Table~\ref{tab:firstmint} in Appendix shows the top 5 of them in terms of the amount of profits.
As we can see, the most profitable one has earned 103.92 Ether from requiring mint fee from buyers on 696 NFTs, whose worth is estimated to be around \$19M.

\subsubsection{Bonus creator fee}
\label{sec:rq2:implicit:creatorfee}
To promote the liquidity of NFTs, a sum of \textit{creator fee} will be charged from the buyer once an NFT is traded on secondary markets.
As a result, some developers will continue to run their projects to gain the confidence and continue to generate revenue from creator fee. Once they believe that the revenue has reached their expectation, they will conduct rug pulls, resulting in financial losses for all holders.

\textbf{Detection.}
To identify this type of rug pulls, we have to extract the money flow of creator fee from secondary markets to contracts of projects. Note that, not all secondary markets explicitly transfer creator fee via transactions. We thus only focus on \texttt{Wywern} of \texttt{OpenSea} (see \S\ref{sec:data:secondary}).
At last, we found that 2,140 projects (28.6\%) have once received creator fee, which is worth over \$12M, as part of its revenue. Table~\ref{tab:top5creatorfee} in Appendix shows the top 5 projects that gain the most creator fee, which sums up to \$4.9M, accounting for 40.8\% of the total.

\subsection{Brief Summary}
\label{sec:rq2:further}
Overall, we have successfully identified additional tricks used in 6,283 (84\%) rug pulled projects, i.e., 667 NFT rug pulls exploited explicit tricks and 5,644 NFT rug pulls used implicit tricks.
Among them, 28 NFT projects take advantage of both ways, and most of them (20) are rug pulled explicitly by hidden URI replacement or unapproved transfer and implicitly by mint fee withdraw. 
This indicates that after withdrawing mint fee as profits, some rug pullers still tend to utilize backdoors to change the URI of NFTs or transfer NFTs out without authorizations, resulting in twice damage to the victims.
Note that we did not observe additional tricks for the remaining 1.2K (16\%) identified rug pulled projects. Through manual verification, we believe that they are ordinary rug pulls that by pumping up the price and selling the NFTs with a high price.

\begin{framed}
\noindent\textbf{Answer to RQ3}
\textit{
Beyond the pump and dump nature of NFT rug pull scams, we have observed that 84\% of the scam projects have exploited at least one kind of tricks to facilitate the delivery of the scam. 
}
\end{framed}
\section{RQ4: Early Warning of NFT Rug Pulls}
\label{sec:predicting}

Our previous exploration has uncovered NFT rug pulls are prevalent in the ecosystem. However, all existing efforts are \textit{postmortem analysis} that rely on reactive methods to flag NFT rug pulls after the scam has happened. It is urgent for our community to raise warnings of such scams at their early stage.
In this section, we seek to explore whether we can identify suspicious NFT projects before rug pull happens based on the initial indicators extracted from transaction and trade events.

\subsection{Dataset and Pre-processing}

For all the labeled 7,487 NFT rug pulls in \S\ref{sec:overall-results:results}, we regard them as the ground truth, denoted as $D_{pos}$.
We further choose the top-1000 projects listed in the \texttt{OpenSea} top collection~\cite{OpenSeatop} as normal projects, denoted as $D_{neg}$. 
Among these 1,000 cases, some of them are not based on Ethereum, thus 933 projects remained.
Considering that the current cases in $D_{neg}$ are generally more active and have higher transaction volumes than those in $D_{pos}$,  we need more cases in the wild to avoid overfitting issue. 
Therefore, to ensure that the new chosen projects are not rug pulls, we sampled 2,000 NFT projects beyond the top-1000 list, still have high liveness of their social media and have no signs of rug pulls, adding them into $D_{neg}$. The first two authors have manually verified that they are definitely not scams by the time of our study.
Finally, we have 2,933 samples in $D_{neg}$ and 7,487 samples in $D_{pos}$.

To train a model for raising early warnings, we need some fine-grained labels on the life-cycle of each project (e.g., 24h before rug pull happens).
Therefore, we have to define some notations on the lifecycle of an NFT project as follows:
\begin{itemize}
\item \textit{$T_{RP}$}: the exact timestamp when the rug pull happens if the given project is rug pulled;
\item \textit{$T_{A}$}: a timestamp ahead of $T_{RP}$, at which we can raise an alarm for suspicious rug pulls;
\item \textit{$P_{FE}$}: a period of time ranging from the launching of the project to $T_{A}$, which is used to extract features;
\end{itemize}
Note that, for cases in $D_{neg}$, both $T_{RP}$ and $T_{A}$ are set as the data of our collection, i.e., $P_{FE}$ would last until the end of data collection. For cases in $D_{pos}$, the first challenge is how to determine $T_{RP}$.
Based on a series of NFT rug pull security reports (collected during the initial 253 ground truth samples), We heuristically design a set of rules, as follows:
\begin{enumerate}
\item If the project has been withdrawn more than once, we will set the $T_{RP}$ as the moment that the largest amount of Ether were withdrawn. 
\item Otherwise, we set the $T_{RP}$ as the moment its corresponding social media accounts last update, if it was not deleted or suspended.
\item Otherwise, if there is a drawdown greater than 0.99 (defined in \S\ref{sec:detectmethod}), we will set the $T_{RP}$ as the moment when the last $p_j$ in the drawdown occurred. This indicates that the investors finally lose confidence on this project.
\item Otherwise, we will set the $T_{RP}$ as the moment of the last market trade record in secondary markets.
\end{enumerate}

If $T_{RP}$ still cannot be determined after these four strategies, in order to promise the accuracy of our prediction model, we will remove the case from $D_{pos}$. Consequently, 5,004 cases remained, and they all have a firm and determined $T_{RP}$. Note that, for the 253 initial collected ground truth samples, our labelled $T_{RP}$ is inline with their original security reports, which indicate the reliability of our labelling method.

Recall that our purpose is to raise warnings of rug pulls before $T_{RP}$. Thus, we decide to use a \textit{time slicing window} to evaluate how early we can accurately predict the rug pull will happen ahead of $T_{RP}$, i.e., the length of $T_{RP} - T_{A}$.
Specifically, because the $T_{RP}$ for these 5K cases of $D_{pos}$ has fixed already, we move $T_{A}$ to extend the length of $T_{RP} - T_{A}$ to give possibilities for investors to transfer assets out as much as possible.
For all cases in either $D_{pos}$ or $D_{neg}$, we set different time slicing windows in different rounds of training, as shown in the first column of Table~\ref{tab:metrics for model}.
After labeling on the life-cycle of cases, we merge $D_{pos}$ and $D_{neg}$, and divide them as 80\% training data and 20\% testing data, using cross validation to evaluate the models.

\subsection{Extracting Features}
\label{sec:rq4:features}
To effectively pinpoint NFT rug pulls in their early stage, we should build an effective classifier.
Therefore, we have extracted a comprehensive set of features that can be divided into three categories, i.e., \textit{time-series}, \textit{on-chain events}, and \textit{secondary markets trade}.
Detailed explanations are fully listed in Table~\ref{tab:feature} in Appendix.

\subsubsection{Time-series Features}
Time-series features are composed of temporal metrics that can be used to flag \textit{oncoming} rug pull projects.
Specifically, we first calculate how long it will take for a project from its launching to its first mint, denoted as $T_{launch\_and\_mint}$. This is inspired by the intuition that rug pullers are in a hurry to obtain profits, who are likely to mint tokens immediately after launching.

Moreover, the distribution of certain activities on the life-cycle of a project is a good indicator. Intuitively, due to the dramatic price pump of rug pulled projects at the early stage, the distribution of various activities will be concentrated to the first or the last part at this moment.
However, for those healthy and innocent ones in $D_{neg}$, the case will be totally different.
Moreover, some trading price related activities, e.g., the timestamp of the floor price occurs, also differ between rug pulled projects and normal ones. 
Thus, we introduce $P_{act}$ to quantify the distribution of activities, which is defined as:
\begin{equation} 
\label{equ:averagetimepoint}
    P_{act}=\frac{\frac{1}{n}\sum_{i=1}^n(T_i-T_{start})}{T_{end}-T_{start}}\nonumber
\end{equation}
, where $n$ is the number of specific type of activities mentioned above, $T_i$ denotes the timestamp of the i-th activity, $T_{start}$ and $T_{end}$ denote the first and last activity of this type, respectively.
For example, for those normal cases, the $P_{swap}$ will be roughly close to 0.5. 
However, it is totally different for cases in $D_{pos}$.
To be specific, due to frequent pump of projects at the beginning of the projects, all swap events will be concentrated at the end of $P_{FE}$, which leads to $P_{swap}$ as 1. Moreover, the pump will also push all top prices occur at the end of $P_{FE}$, while it tends to be random for normal projects.
In a nutshell, $P_{act}$ can measure the degree of concentration of distribution of a certain activity.
We have extracted 9 features in total, including certain type of events and prices, which are shown in Table~\ref{tab:feature} in Appendix.

%\noindent
\subsubsection{Features of On-chain Events}
On-chain events are emitted by on-chain token transfers, which can reflect the liveness of the corresponding projects. Thus, we pay attention to events from two perspectives, i.e., the total number (denoted as $N_{event}$), and the number of involved participants (counted by addresses, denoted as $A_{event}$).
For example, as discussed in \S\ref{sec:overall-results:results}, $N_{mint}$ is often higher than $N_{swap}$ for rug pulled projects because of the inborn poor liquidity of rug pulls.
But for non-rug-pull projects, it is always the opposite. In addition, $A_{mint}$ refers to how many accounts are participated in minting, which could be extremely low, e.g., NFTs are all minted to a middleman (see \S\ref{sec:middleman}).
Furthermore, we think the ratio of certain type of events to all emitted events can also reflect some characteristics. To depict the above features, we thus defined another two features, denoted as $RN_{event}$ and $RA_{event}$.
For example, swap events for rug pulled projects only account for a small part of all emitted events, leading to an extremely low $RN_{swap}$.
Totally, we have extracted 15 features, related to three types of events (see Table~\ref{tab:feature} in Appendix).

\subsubsection{Features of Secondary Market Trades}
Similarly, the liveness of projects in secondary markets can also be reflected by features extracted from initiated trades. Due to rug pulled projects would always manipulate the mechanisms of secondary markets, they will definitely display different traits, which can be further utilized in $P_{FE}$ to distinguish cases.
Specifically, we mainly focus on three metrics, i.e., $N$, $U$, and $V$, 
referring to the total number, involved users, and price or trading volumes. For example, $N_{trade}$ stands for the total number of trades, and $V_{average\_price}$ refers to the average price of all trades.
Moreover, we still think the ratio can reflect some characteristics that can be utilized. 
We add a prefix $R$ for those three metrics.
For example, $RU_{highest\_24h}$ means the ratio of the users involved in one day that has the highest trading volume to all involved users in history.
In total, we have extracted 31 features shown in Table~\ref{tab:feature} in Appendix.
\begin{table}[t]
\centering
\caption{Metrics of different models and $T_{RP}-T_{A}$, where $P$, $R$, $F1$ refer to \textit{precision}, \textit{recall}, and \textit{F1 score}, respectively.}
\label{tab:metrics for model}
\resizebox{\columnwidth}{!}{%
\begin{tabular}{cccccccccc}
\toprule
\multirow{2}{*}{\begin{tabular}[c]{@{}c@{}}$T_{RP}-T_{A}$\\(hour)\end{tabular}} & \multicolumn{3}{c}{\textbf{Logistic Regression}} & \multicolumn{3}{c}{\textbf{SVM}}      & \multicolumn{3}{c}{\textbf{Random Forest}} \\
                       & $P$     & $R$     & $F1$    & $P$ & $R$ & $F1$ & $P$   & $R$  & $F1$  \\ \midrule
0  & 0.95 & 0.98 & 0.96 & 0.95 & 0.98 & 0.96 & 0.85 & 0.79 & 0.87 \\
1  & 0.93 & 0.98 & 0.94 & 0.93 & 0.98 & 0.95 & 0.86 & 0.78 & 0.87 \\
2  & 0.94 & 0.98 & 0.95 & 0.94 & 0.99 & 0.95 & 0.86 & 0.79 & 0.88 \\
4  & 0.94 & 0.97 & 0.95 & 0.94 & 0.97 & 0.95 & 0.86 & 0.78 & 0.87 \\
8  & 0.91 & 0.96 & 0.93 & 0.91 & 0.97 & 0.93 & 0.83 & 0.76 & 0.84 \\
12 & 0.92 & 0.98 & 0.93 & 0.92 & 0.98 & 0.93 & 0.85 & 0.75 & 0.85 \\
16 & 0.88 & 0.95 & 0.90  & 0.88 & 0.96 & 0.90  & 0.85 & 0.76 & 0.84 \\
24 & 0.91 & 0.98 & 0.92 & 0.92 & 0.98 & 0.93 & 0.86 & 0.75 & 0.85 \\
36 & 0.90  & 0.94 & 0.90  & 0.90  & 0.94 & 0.91 & 0.84 & 0.7  & 0.82 \\ 
48  & 0.94 & 0.96 & 0.95 & 0.94 & 0.96 & 0.95 & 0.85 & 0.77 & 0.86 \\ 
60  & 0.93 & 0.95 & 0.94 & 0.92 & 0.96 & 0.93 & 0.84 & 0.74 & 0.84 \\
72  & 0.91 & 0.95 & 0.92 & 0.91 & 0.95 & 0.92 & 0.85 & 0.75 & 0.85 \\
84  & 0.92 & 0.96 & 0.93 & 0.92 & 0.98 & 0.93 & 0.86 & 0.77 & 0.86 \\
96  & 0.92 & 0.97 & 0.93 & 0.92 & 0.97 & 0.93 & 0.84 & 0.73 & 0.83 \\ 
\bottomrule
\end{tabular}%
}
\end{table}

\subsection{Model Training \& Predicting Result} 
\label{sec:training-predicting}
We adopt three widely-adopted machine learning algorithms, i.e., logistic regression~\cite{kleinbaum2002logistic}, SVM~\cite{cherkassky2004practical}, and random forest~\cite{breiman2001random}.
Three metrics are calculated to quantify the result, i.e., \textit{precision}, \textit{recall}, and \textit{F1 score}.
Table~\ref{tab:metrics for model} shows the results of metrics on different time slicing windows, i.e., $T_{RP}-T_{A}$.

Generally speaking, compared to random forest algorithm, both logistic regression and SVM can obtain a quite good result (e.g., roughly 90\% precision, recall, and F1) when the slicing window is set up to 96 hours (4 days earlier before the scam happends).
This may be due to the fact that random forest only selects a part of features for each round of the model training process. If the features are split, they may not effectively distinguish cases from $D_{pos}$ and $D_{neg}$.

\subsection{Real-time Monitoring}

\subsubsection{Real-world Settings}
To apply our method in practice, we have implemented a prototype system and launched it since Nov. 1st, 2022, and conducted a prediction once a day.
Specifically, we will set the $T_{RP}-T_A$ of different time slicing windows (see Table~\ref{tab:metrics for model}), and predict once a day at midnight if the given project will be rug pulled in the next following day. All features will be extracted and calculated from the birth of each project to the day of the prediction.
As discussed in \S\ref{sec:training-predicting}, both logistics regression and SVM can effectively predict rug pulls. Thus if any of these two models raises alarms, we will label it as an oncoming rug pull project.

\subsubsection{Result}
Our system raises alarms everyday, ranging from 250 to 1,500 reports, since its deployment. By the time of Apr. 1st, 2023 (151 days), the number of alarms is 950 everyday on average, and we have predicted 7,821 projects that have the possibility to be rug pulled in total after removing duplication.
Although it is hard for us to accurately verify whether these projects will be rug pulled in the near future, we have observed some additional evidences that can support our prediction. 

\textit{First, their social network activities show strong hints.}
For example, we have alerted 5,833 NFT projects from November 2022 to January 2023, and we have collected the social media information for 41.3\% of them.
When we investigate these projects in April 2023 (i.e., three months later), we observe that 68.7\% of them are abnormal. To be specific, 108 projects have been suspended, 30 projects have deleted all their posts, 835 projects have not updated for at least one month and 684 have been closed by the platform. 
These are strong indicators that the rug pulls have been taken place.
\textit{Second, we have reported some popular NFT projects that we identified to a leading blockchain security company (anonymized) for confirmation.} 
For example, \texttt{CheckPunks}\footnote{\href{https://etherscan.io/address/0xf7af6dd6c36fc42eb802b33d48ea3d395803c8ea}{https://etherscan.io/address/0xf7af6d..}} is a project that counterfeits the famous \texttt{CryptoPunks}~\cite{cryptopunks}, holding total volume of 258 Ether and over 10,000 trades in secondary markets. 
Our system raises alarm of this project since Jan. 24th, 2023. 
However, \texttt{ChecksPunks} is still active in \texttt{OpenSea} in the early February~\cite{checkpunksopenseaanalysis}. Since March, its official Twitter account has been abandoned~\cite{checkpunkstwitter}. 
In addition, Risk Radar Chart from \texttt{MetaDock} also labelled the top-tier market risk level for \texttt{CheckPunks}~\cite{checkpunksopensea}, which further confirms our prediction. 
Our system alarms it at the very early stage which can prevent more damage to the investors and the NFT community.
\textit{Third, we have manually investigated our prediction results to see if they have actually been rug pulled later and we further analyzed the longest time sliding window that can work in the wild.} 
We sampled 200 projects for manual verification, based on their transaction behaviors (e.g., drain the money out).
We observed that over 90\% of them have been rug pulled afterwards by the time of this writing. 
For example, \texttt{8 Bit Heroes}\footnote{\href{https://etherscan.io/address/0x3f56312caa4edff778f37173d200b60ef0ec29be}{https://etherscan.io/address/0x3f5631...}} is a collection of 4,808 NFTs, which launched on Jan 2nd, 2023. 
However, it has withdrawn all the Ether in the contract until Jan. 21th, 2023\footnote{\href{https://etherscan.io/tx/0x83f0748292524de4585c1798a154dcac2f19aabd3add446328c81b9f2fb8c3d0}{https://etherscan.io/tx/0x83f074...}}, after which it stopped updating.
However, we start to raise alarms of this project since Jan. 8th, 2023, 13 days before its rug pull.

\begin{framed}
\noindent\textbf{Answer to RQ4}
\textit{
We can indeed identify suspicious NFT projects before rug pull happens based on the
initial indicators extracted from transaction and trade events. Our system can work as a whistle blower that pinpoints rug pull scams timely, thus mitigating the impacts.
}
\end{framed}

\label{sec:forcastnewrugpulls}
\section{Discussion}
\subsection{Lessons Learned}
\label{sec:discussion}
Our work reveals that rug pull events are rampant in the NFT ecosystem.
Therefore, we summarize some suggestions for investors, developers and secondary markets managers.

\noindent
\textbf{For investors.}
Several best practices can be given for NFT investors.
First, only open-source projects that are audited by prestigious security companies are acceptable. Otherwise, explicit rug pulls may occur due to backdoors in smart contracts (see \S\ref{sec:rq2:explicit}).
Second, pay special attentions to the NFT projects that require a sum of mint fees, as withdrawing mint fee is one of the mainstream profiting way for rug pullers.
Third, do not be fooled by the seemingly prosperity of NFT projects.
If a popular NFT can be bought by a lower price than market price, pay attention to examine if it is a ``mirage''.

\noindent
\textbf{For developers.}
For a better NFT ecosystem, developers should avoid rug pulls happen, which can be generally divided in twofold.
On the one hand, developers should strictly follow the best practices of implementing ERC standards. For example, use \texttt{totalSupply} instead of other self-defined variables to limit the circulation of available NFTs. Developers should also open-source the implementation or even ask for code auditing and bug bounty to eliminate investors concern. 
On the other hand, developers should pay attention to behaviors that may result in misunderstanding of investors, e.g., requiring a bunch of mint fees, issue all NFTs to a single account.
In general, building a project's reputation and gaining users' confidence heavily require efforts from developers.

\noindent
\textbf{For secondary markets managers.}
First, all circulated NFTs should be critically reviewed to avoid counterfeit which brings in financial losses for both trading platforms and holders.
Second, creator fee is a huge part of profits for rug pullers, which urges the disclosure and traceability of all trading transactions to track attackers if rug pulls happened.
Last but not least, the predicting method proposed in this paper is proven to be effective and efficient. It is reasonable and practical to integrate such predicting methods into secondary markets to raise alarms for holders in advance.

\subsection{Threats of Validity}
\label{sec:limitation}
% \jt{I revised the section.}
Our study carries certain limitations.
First, our data of secondary markets is incomplete. As we discussed in \S\ref{sec:data:secondary}, we only collected trades from the top three secondary markets in terms of trading volumes. 
However, these three have accounted for over 83\% of total trading volumes in Ethereum for \texttt{ERC-721} and \texttt{ERC-1155} NFTs, which means including other markets will not influence the final results significantly.
Second, our rule-based approach in RQ2 is quite straightforward, which heavily depend on the symptoms we summarized from the pilot study.
However, to get a reliable results, we make them quite conservative for both rules and our detector. To this end, we can guarantee that we can identify the lower bound of rug pulled NFT projects.
Third, there might be some advanced tricks used by the rug pull scams we did not cover in this paper, as manual efforts are widely used in this work. Even so, to the best of our knowledge, we have conducted the most comprehensive study on NFT rug pull so far.

\section{related work}

\noindent
\textbf{NFT measurement.}
Following the surge in popularity of NFTs in 2021, several researchers have focused on this area~\cite{wang2021non,bao2021recent,bao2022non,chohan2021non,dowling2022fertile,borri2022economics,kugler2021non,chandra2022non,white2022characterizing,gupta2022identifying,bhujel2022survey,la2022nft,serneels2023detecting,tariq2022suspicious}.
In 2021, Wang et al.~\cite{wang2021non} introduced the NFT ecosystem as a first step. Kugler et al.~\cite{kugler2021non} proposed the use of non-fungible tokens and measured their economic impact. In 2022, White et al.~\cite{white2022characterizing} conducted a study on OpenSea, and found that despite sparsity in the network, communities of users are forming and power users tend to congregate in these structures. In 2023, Roy et al.~\cite{roy2023demystifying} used machine learning to detect NFT phishing. Gupta et al.~\cite{gupta2022identifying} conducted a security survey of the NFT ecosystem and identified various security issues. In addition, Von et al.~\cite{von2022nft} used different methods to detect wash trading behavior in the NFT ecosystem on Ethereum.
However, as the most prominent type of scam, NFT rug pulls have not been systematically explored. Our research serves as the first to detect and analyze NFT rug pulls, which is of significant importance to stakeholders in the NFT community.

\noindent
\textbf{Cryptocurrency rug pull.}
In 2021, Xia et al.~\cite{xia2021trade} employed machine learning methods to identify scam tokens in Uniswap, a decentralized exchange of DeFi. Among the scam schemes, rug pulls were discussed in their work. In 2022, Mazorra et al.~\cite{mazorra2022not} introduced the environment of ERC-20 tokens (fungible tokens) and two types of confirmed rug pulls. They used various methods, including activity-based and machine learning with hyperparameter optimization, to detect rug pulls. Scharfman et al.~\cite{scharfman2022decentralized} discussed DeFi case studies, including rug pulls, pump and dump scams, and regulatory actions involving DeFi. Cernera et al.~\cite{cernera2022token} discussed three types of malicious behavior in Binance, including rug pulls.
Our work builds upon previous research on rug pulls but is specific to NFT projects, as the characteristics of NFTs are distinct from those of previous work. We propose a method to detect NFT rug pulls and raise alarms ahead of the scam happens, which is different with all existing work.

\section{conclusion}
This paper presents the first comprehensive study of NFT rug pulls. By summarizing the key symptoms of rug pull scams, we have formulated a list of concrete rules to flag rug pull projects in the NFT ecosystem, and curated a list of 7,487 rug pull projects, by far the largest dataset of NFT rug pulls. We further designed checkers to uncover diverse sophisticated tricks used in them. To further impede the expansion of the scam, we further design a prediction model to proactively identify the potential rug pull projects in an early stage ahead of the scam happens. This paper presents the first solution to detect, mitigate and even prevent NFT rug pulls.

\bibliographystyle{IEEEtran}

\centering
\appendix
\setcounter{table}{2}

\begin{table*}[htbp]
\centering
\caption{Features are used in the NFT rug pull classifier (see \S\ref{sec:predicting}). The feature that cannot be calculated (e.g., due to the lack of data) will be set as -1.
}
\resizebox{0.9\textwidth}{!}{
\begin{tabular}{|c|cc|}
\hline
                              & Feature               & Description                                                                                         \\\hline
\multirow{9}{*}{\rotatebox{90}{Time-series}}         
                              & $T_{launch\_and\_mint}$       & Time period between the launch and the first mint event         \\\cline{2-3}
                              & $P_{transfer}$       & Average timepoint of each transfer        \\\cline{2-3}
                              & $P_{mint}$        & Average timepoint of each mint event       \\\cline{2-3}
                                & $P_{swap}$        & Average timepoint of each swap event    \\\cline{2-3}
                                  & $P_{burn}$        & Average timepoint of each burn event    \\\cline{2-3}
                              & $P_{trade}$        & Average timepoint of each trade            \\\cline{2-3}
                              & $P_{top\_price}$        & The timepoint of the trade with highest price           \\\cline{2-3}
                              & $P_{floor\_price}$        & The timepoint of the trade with the lowest price       \\\cline{2-3}
                              & $P_{highest\_24h\_trade}$        & Average timepoint of the trade with the highest trade events in 24 hours      
                              \\\cline{1-3}
\multirow{14}{*}{\rotatebox{90}{Token Transfer Logs}}     & $N_{Transfer}$              & Total times of transfer event
                                                    \\\cline{2-3}
                              & $N_{mint}$             & The number of mint event
                                                  \\\cline{2-3}
                              & $N_{swap}$            & The number of swap event
                                              \\\cline{2-3}
                              & $N_{burn}$            & The number of burn event
                                              \\\cline{2-3}
                 %              & $RN_{transfer\_tx}$          & The ratio between the transfer events and transactions
                 % \\\cline{2-3}
                              & $RN_{mint\_transfer}$        & The ratio between the mint events and transfer events
                     \\\cline{2-3}
                              & $RN_{swap\_transfer}$        & The ratio between the swap events and transfer events
             \\\cline{2-3}
                              & $RN_{burn\_transfer}$            & The ratio between the burn events and transfer events
                                              \\\cline{2-3}
                                                
                              & $A_{all}$          & The number of addresses that have participated in transfer events         \\\cline{2-3}
                              & $A_{mint}$            & The number of addresses that have participated in mint events        \\\cline{2-3}
                              & $A_{swap}$          & The number of addresses that have participated in swap events      \\\cline{2-3}
                              & $A_{burn}$        & The number of addresses that have participated in burn events     \\\cline{2-3}
                              & $RA_{mint\_all}$           & The ratio between the mint events and transfer events        \\\cline{2-3}
                                 & $RA_{swap\_all}$             &The ratio between the swap events and transfer events                \\\cline{2-3}
                              & $RA_{burn\_all}$       & The ratio between the burn events and transfer events                                                                       \\\cline{1-3}
\multirow{30}{*}{\rotatebox{90}{Secondary Market Trades}} & $N_{trade}$             & Total times of trade\\\cline{2-3} &
                                $V_{volume}$              & The history volume of trade\\\cline{2-3}
                                
                              & $V_{average\_price}$         & The average price of trades\\\cline{2-3}
                              & $N_{beyond\_average}$         & The number of trades whose price is beyond the average price\\\cline{2-3}
                              & $N_{below\_average}$         & The number of trade whose price is below the average price \\\cline{2-3}
                              & $RN_{beyond\_average}$         & The ratio between the ``beyond average'' trades and all trades\\\cline{2-3}
                              & $RN_{below\_average}$         & The ratio between the ``below average'' trades and all trades\\\cline{2-3}
                              
                              & $V_{top\_price}$         & The highest price of the trades.\\\cline{2-3}
                              & $V_{floor\_price}$         & The lowest price of the trades. \\\cline{2-3}
                              
                              & $U_{all}$         & Total users that have participated in trade \\\cline{2-3}
                              & $U_{buyer}$         &Total users who are buyers\\\cline{2-3}
                              & $U_{seller}$         & Total users who are sellers\\\cline{2-3}
                              & $RU_{buyer\_all}$            &The ratio between the buyers and all users\\\cline{2-3}
                              & $RU_{seller\_all}$         &  The ratio between the sellers and all users \\\cline{2-3}
                              
                              & $N_{highest\_24h\_trade}$         &  The number of trades of day that have the highest amount of trades events \\\cline{2-3}
                              & $RN_{highest\_24h\_trade}$         &  The ratio between $N_{highest\_24h\_trade}$ and all trade events \\\cline{2-3}
                              & $V_{highest\_24h\_volume}$         &  The history volume of the day that have highest trades events \\\cline{2-3}
                              & $RV_{highest\_24h\_volume}$         &  The ratio between $D_{highest\_24h\_volume}$ and total volume \\\cline{2-3}
                              & $V_{highest\_24h\_average\_price}$         &  The average price of the day that have highest trades events \\\cline{2-3}
                              & $RV_{highest\_24h\_average\_price}$         &  The ratio between $V_{highest\_24h\_average\_price}$ and average price\\\cline{2-3}
                              & $U_{highest\_24h\_user}$         &  The number of users of the day that have highest trades events \\\cline{2-3}
                              & $RU_{highest\_24h\_user}$         &  The ratio between $U_{highest\_24h\_user}$ and total number of  users \\\cline{2-3}
                              & $N_{recent\_24h\_trade}$         &  The number of trade events of the recent day\\\cline{2-3}
                              & $RN_{recent\_24h\_trade}$         &  The ratio between $N_{recent\_24h\_trade}$ and all trade events\\\cline{2-3}
                              & $V_{recent\_24h\_volume}$         &  The history volume of the recent day \\\cline{2-3}
                              & $RV_{recent\_24h\_volume}$         &  The ratio between $V_{recent\_24h\_volume}$ and total volume \\\cline{2-3}
                              & $V_{recent\_24h\_average\_price}$         &  The average price of the recent day \\\cline{2-3}
                              & $RV_{recent\_24h\_average\_price}$         &  The ratio between $V_{recent\_24h\_average\_price}$ and average price \\\cline{2-3}
                              & $U_{recent\_24h\_user}$         &  The number of users of the recent day \\\cline{2-3}
                              & $RU_{recent\_24h\_user}$         &  The ratio between $U_{recent\_24h\_user}$ and total number of users
                              \\\cline{1-3}
\end{tabular}
}
\label{tab:feature}
\end{table*}

\begin{table*}[h]
\centering
\caption{Top 5 profitable NFT projects rug pulled by mint fee withdraw.}
\label{tab:mintfee-withdraw}
\begin{tabular}{@{}ccc@{}}
\toprule
\textbf{Project}    & \textbf{Contract Address}                    & \textbf{\# Ether (USD)} \\ \midrule
Apes In Space    & 0x7a3b97a7400e44dadd929431a3640e4fc47daebd & 2,645.82 (10.93M)       \\
Fat Ape Club     & 0xf3114dd5c5b50a573e66596563d15a630ed359b4 & 2,573.00 (11.70M)       \\
Bored Bunny      & 0x9372b371196751dd2f603729ae8d8014bbeb07f6 & 1,991.00 (7.08M)        \\
MURI             & 0x4b61413d4392c806e6d0ff5ee91e6073c21d6430 & 1,872.60 (4.92M)        \\ 
HULLYUniverse             & 0xb8b6cb37c0968f72c6d37dc3074c80ad73521024 & 1398.90 (3.92M)        \\ 
\bottomrule
\end{tabular}%
\end{table*}

\begin{table*}[h]
\centering
\caption{Top 5 NFT projects in terms of participating wash trading.}
\label{tab:washtrading}
\begin{tabular}{@{}ccc@{}}
\toprule
\textbf{Project}     & \textbf{\begin{tabular}[c]{@{}c@{}}Addresses of Suspicious\\ Seller \& Buyer\end{tabular}} & \textbf{\begin{tabular}[c]{@{}c@{}}\# Wash Trading Trades\\ (\$ of History Volumes)\end{tabular}} \\ \midrule
\multirow{2}{*}{Audioglyphs} & 0x37929647c6bab7033f8d902a31a3afbae3767e69                                                 & \multirow{2}{*}{\begin{tabular}[c]{@{}c@{}}735\\ (377.8M)\end{tabular}}                                  \\
                     & 0x72c3dfe90b733c236b0e8c200dc71eea123c3dca                                                 &                                                                                                          \\
\multirow{2}{*}{CATGIRL ACADEMIA} & 0x7101075a76296b60ec2d8571ae2aae301b2caa21                                                 & \multirow{2}{*}{\begin{tabular}[c]{@{}c@{}}416\\ (339.9M)\end{tabular}}                                  \\
                     & 0xcc1aa6d6d0e9e8876b7f41f384a155e0774ae7b6                                                 &                                                                                                          \\
\multirow{2}{*}{CryptoPhunksV2} & 0xa6e3bd38f3399037fa75088516a3935bbb08ad16                                                 & \multirow{2}{*}{\begin{tabular}[c]{@{}c@{}}140\\ (275.8M)\end{tabular}}                                  \\
                     & 0x44e37065db06958e6d84f88d688eeb5661d6fa7d                                                 &                                                                                                          \\
\multirow{2}{*}{Last Boy Standing} & 0x0a26fbdfe91e0aa6bd54547dec23f1bfe31874d1                                                 & \multirow{2}{*}{\begin{tabular}[c]{@{}c@{}}197\\ (38.9M)\end{tabular}}                                   \\
                     & 0x2b82027683c58cfb817bab52f3c4f10ff6fbbd92                                                 &                                                                                                          \\
\multirow{2}{*}{Metasaurs} & 0xe2bf62b749450d13a03212118fee055134bf8211                                                 & \multirow{2}{*}{\begin{tabular}[c]{@{}c@{}}20\\ (3.0M)\end{tabular}}                                     \\
                     & 0x63d9d24a199e74c32e324c199139e3df9ced13f4                                                 &                                                                                                          \\ \bottomrule
\end{tabular}%
\end{table*}

\begin{table*}[h]
\centering
\caption{Top 5 NFT projects gain the largest profit via middleman reselling.}
\label{tab:firstmint}
\begin{tabular}{@{}lccc@{}}
\toprule
\multicolumn{1}{c}{\textbf{Project}} & \textbf{Middleman Address}                & \textbf{\# mint tokens} & \textbf{Ether (USD)}   \\ \midrule
\multicolumn{1}{c}{CypherHumans}              & 0x6e40ea6202d5bc2ace21bc904c9c772c484320a1 & 696               & 103.92 (188K) \\
\multicolumn{1}{c}{Muttniks}                  & 0xa00f56b263d3c3e016c33d9b31791b625d90ae3b & 1080              & 64.20 (116K)   \\
\multicolumn{1}{c}{DigDragonz}                & 0x0ce353f8bca317024e4ae6b87a0e14ca0377f476  & 944               & 46.35 (83K)   \\
\multicolumn{1}{c}{DigDragonzReborn}          & 0xcd6d7e5a31cb3cf43734398e6506d5422072a172 & 1507              & 27.45 (50K)   \\
\multicolumn{1}{c}{DRM1}                      & 0x171ab540b9cb730626db91f648e2b09eb5363484 & 101               & 19.99 (36K)   \\ \bottomrule
\end{tabular}%
\end{table*}

\begin{table*}[h]
\centering
\caption{Top 5 NFT rug pull projects that earn most from bonus creator fee.}
\label{tab:top5creatorfee}
\begin{tabular}{ccc}
\hline
\textbf{Project}      & \textbf{Contract Address}                            & \textbf{Creator Fee (USD)} \\ \hline
MURI              & 0x4b61413d4392c806e6d0ff5ee91e6073c21d6430 & 1.486M                  \\
SkuxxVerse Pass   & 0x19350eb381ab2f88d274e740bd062ab5ff15542e & 1.198M                  \\
hausphases        & 0x5be99338289909d6dbbc57bb791140ef85ccbcab & 0.958M                   \\
Beyond Earth Land & 0x28c6ea3f9cf9bc1a07a828fce1e7783261691b49 & 0.797M                   \\
Moonbirds2        & 0xdb7b094fdc04f51560a03a99f747044951b73727 & 0.490M                   \\ \hline
\end{tabular}%
\end{table*}

\end{document}